\documentclass{aa}
\usepackage[varg]{txfonts}
\usepackage{graphicx}
\usepackage{natbib}

\newcommand{\kms}{$\rm km\,s^{-1}$}

\newcommand{\hi}{\ion{H}{I}}

\begin{document}
\title{A new catalog of  \hi\, supershell candidates in the outer part of the Galaxy}
\author{L. A. Suad\inst{1}, C. F. Caiafa\inst{1,2}, E. M. Arnal\inst{1,3}, \and S. Cichowolski\inst{4}
}

\institute{Instituto Argentino de Radioastronom\'{\i}a (IAR), CC 5,
  1894, Villa Elisa, Argentina. 
  \and Facultad de Ingenier\'{i}a, Universidad de Buenos Aires, C.A.B.A., Argentina.
  \and Facultad de Ciencias Astron\'omicas y Geof\'{\i}sicas, Universidad Nacional de La Plata, La Plata, Argentina.
  \and   Instituto de Astronom\'{i}a y F\'{i}sica del Espacio (IAFE), Ciudad Universitaria, C.A.B.A, Argentina.
}

\date{received  /accepted}

\abstract {}{The main goal of this work is to a have a new neutral hydrogen (\hi) supershell candidates catalog to analyze their spatial distribution in the Galaxy and to carry out a statistical study of their main properties.}{This catalog was carried out making use of the Leiden-Argentine-Bonn (LAB) survey. The  supershell candidates were identified using a combination of two techniques:  a visual inspection plus an automatic searching algorithm. Our automatic algorithm is able to detect both closed and open structures.}{A total of 566 supershell candidates were identified. Most of them (347) are located in the second Galactic quadrant, while 219 were found in the third one. About $98 \, \%$ of a subset of 190 structures (used to derive the statistical properties of the supershell candidates) are elliptical  with a mean  weighted eccentricity of $0.8 \pm 0.1$, and $\sim 70 \,\%$  have their major axes parallel to the Galactic plane. The weighted mean value of the effective radius of the structures is $\sim$ 160 pc. Owing to the ability of our automatic algorithm to detect open structures, we have also identified some ``galactic chimney'' candidates. We find an asymmetry between the second and third Galactic quadrants in the sense that in the second one we detect structures as far as 32 kpc, while for the 3rd one the farthest structure  is detected at 17 kpc. The supershell surface density in the solar neighborhood is $\sim$ 8 kpc$^{-2}$, and decreases as we move farther away form the Galactic center. We have also compared our catalog with those  by other authors.}{}


\titlerunning{\hi\, supershell candidates catalog in the outer part of the Galaxy}

\authorrunning{L. A. Suad et al.}
\maketitle

\section{Introduction}

When viewed in the neutral hydrogen (\hi) line emission ($\lambda = 21$ cm), the interstellar medium (ISM) reveals a complex structuring that manifests itself by the presence of a plethora of structures such as shells, supershells, filaments, arcs, cavities, worms, and loops.
In particular, \hi\, shells and supershells are detected in a given radial velocity range as voids in the \hi\, emission distribution that are  surrounded completely, or partially, by walls of enhanced \hi\, emission.

The physical processes usually invoked  to explain the formation of  \hi\, shells and supershells are the combined action of  strong  winds from massive stars upon the surrounding interstellar medium, and their ultimate explosion as supernovae.  Other alternative mechanisms have also been proposed to explain their creation, such as gamma-ray bursts \citep{efr98,loe98} and the infall of high velocity clouds \citep{ten88}. 

Several catalogs of shells and supershells have already been constructed  with a variety of different techniques such as visual identification \citep{hei79, hei84, mcc02} and automatic identification algorithms \citep{ehl05,dai07,ehl13}. 

Based on a visual inspection of photographic representations of the \hi\, emission distribution derived from the  \hi\, survey of \citet{wea73}, a first catalog of \hi\, shells and supershells was constructed by \citet{hei79}, who defines as supershells to those structures requiring at least $3 \times 10^{52}$ erg for their creation. A total of  63 structures were identified. Later, using the 21 cm database of \cite{hei74} for $\vert b \vert > 10^\circ$, \citet{hu81} discovered 50 new shells. In a subsequent work, \citet{hei84} combined the  surveys of  \cite{wea73} and \cite{hei74} in order to eliminate the boundary problems at $\vert b \vert = 10^\circ$, in this way finding a total of 42 new structures. Afterwards, \citet{mcc02} report the discovery of 19 new \hi\, shells, using the Southern Galactic Plane Survey (SGPS) \citep{mcc01b}.

On the other hand, applying  a fully automatic method  to the \hi\,  Leiden-Dwingeloo survey \citep{har97}, \cite{ehl05} searched for regions having  a local minimum  in the \hi\, emissivity  that were completely encircled by regions of higher emissivity. Later on, the same authors \citep{ehl13} used the Leiden-Argentine-Bonn (LAB) survey \citep{kal05} to conduct an all-sky search for \hi\, shells, employing an automatic procedure that is slightly different from the one described in \cite{ehl05}.
Using  an artificial neural network and the \hi\, database of the Canadian Galactic Plane Survey (CGPS) \citep{tay03},  \cite{dai07} identified a large number of small expanding \hi\, shells  in the Perseus arm. 

Bearing in mind that different identification techniques applied to the same database \citep[see][]{ehl05,ehl13} provide different results and that quite different structures are identified in the same region of the sky when different  identification criteria and databases are used  (see feature GSH\, 263+00+47 of \citeauthor{mcc02} \citeyear{mcc02} and GS\,263-02+45 of \citeauthor{arn07} \citeyear{arn07}), we believe that it is worth making the effort to unveil  the presence of these structures (shells and supershells) by using a relatively different approach for their identification.

In particular, in this paper we deal solely  with the identification of the so-called \hi\, supershells  located in the outer part of the Galaxy by using a novel procedure that combines both visual and automatic algorithms of identification. The constraint of only analyzing the second and third galactic quadrants stems from the fact that toward this part of the Galaxy the radial velocity-distance relationship only has a single value, a fact that strongly simplifies  determination  of the physical sizes of the detected structures. 

The paper is organized as follows. In Section 2 we briefly describe the database used in constructing the catalog  of \hi\, supershell candidates, while in Section 3 the identification techniques are described in some detail. Section 4 describes the selection effects of our catalog. The statistical properties of the supershell candidates are  presented  in Section 5, and a comparison with previous catalogs is made in Section 6.

\section{Observations}  

Neutral hydrogen (\hi) data were retrieved from the
Leiden-Argentine-Bonn (LAB) survey \citep{kal05} to explore the \hi\, emission distribution. This database, well
suited to a study of large scale structures, has an angular
resolution of 34\arcmin, a velocity resolution of 1.3 \kms,  and a channel separation of 1.03 \kms, and it covers
the velocity range from $-400$ to $+450$ \kms. The entire database has
been corrected for stray radiation \citep{kal05}.

\section{catalog of \hi\, supershell candidates}

\subsection{Selection criteria}\label{selec-criteria}

In this work a  given \hi\, structure will be cataloged as a candidate object to be a supershell, if it simultaneously fulfills the following set of conditions:

\begin{itemize}

\item \textbf{a)}  It must have,  in a given velocity range, a well defined low brightness temperature region, which is surrounded, partially or completely, by a ridge of higher \hi\, emissivity.
\item \textbf{b)} The  \hi\, minimum must be observable in at least five
 consecutive velocity channels. 
\item  \textbf{c)} The overall structure (minimum + surrounding ridge) must have a minimum angular size of $2^\circ$. 
\item  \textbf{d)} At the kinematic distance of the \hi\, structure, its linear size  must be larger than 200 pc.
\end{itemize}

Condition \textbf{b)} is set in order to  assure the persistence of the structure  along a minimum velocity  range of $\sim 6$ \kms.
With this criterion, on the one hand, we seek to avoid picking up structures  that may arise from the turbulent nature of the ISM, and  on the other, we  keep the chance of detecting slowly expanding structures.
Condition  \textbf{c)} is related to the angular resolution of the HI survey  we are working with, and it implies that the feature is, angularly speaking, fully resolved.
Finally, criteria \textbf{d)} is imposed to ensure identification of only large scale features.

\subsection{catalog elaboration}\label{elab}

To construct a catalog of structures satisfying the above criteria, a  four-stage procedure was followed, namely:

\begin{enumerate}

\item  A visual search of the supershell candidates. 

\item A \textit{learning phase} of the automatic detection algorithm.

\item Running the automatic detection  algorithm in a blind way.

\item A new visual inspection of the structures found in the previous step.

\end{enumerate}

\noindent In the following sections we describe each step in some detail.

To reach a balance between data-cube size and the computational resources needed to run the algorithm on a reasonable timescale, the individual \hi\, data cubes used in this work were $(\Delta l, \Delta b) = (50^\circ \times 60^\circ$) in size. In Galactic latitude the region $-50^\circ \le b \le 50^\circ$  was covered, while our search covered $80^\circ \le l \le 290^\circ$ in Galactic longitude. Every individual \hi\, cube overlaps $10^\circ$ in Galactic longitude and  $20^\circ$ in Galactic latitude, with the neighboring one.

\subsubsection{Visual search}

The main goal of this step is to identify by eye the most conspicuous structures  likely to be classified as \hi\, supershells. This method, {\it ``due to the ability of the human eye to combine disconnected features into a single shape''} \citep{ehl05} is  very efficient at unveiling either complete or incomplete large angular size features.
To this end,  a visual inspection of the \hi\, data cubes covering the Galactic longitude ranges of $90^\circ  \leqslant l \leqslant 165^\circ$ and $195^\circ  \leqslant l \leqslant 270^\circ$ for $\mid b \mid \, \leqslant 50^\circ$ was performed. The region between $165^\circ  \leqslant l \leqslant 195^\circ$ around the Galactic anti-center was not surveyed because along these lines of sight galactic rotation models provide very unreliable kinematic distances.

The visual search was made using the software KARMA\footnote{This is a visualization software package developed by Richard Gooch, formerly of the Australia Telescope National Facility (ATNF)}, which is a visualization software package that contains several visualization and analysis tools. In this work we have used the module \textit{kvis} to visualize the \hi\, data cubes. Since with \textit{kvis} the position-position images can be seen as  movies of the data cube, it allows us to detect the continuity of a structure at different velocity channels.

It is important to mention that the visual search was independently made by three of us (Suad, L.A.; Cichowolski, S.; and Arnal, E.M.). The identifications made by each of us were compared with the others. Only when a given structure was detected by at least two of us  was it incorporated in the listing of the visual catalog.
In this way, a total of 149 features were listed. Using a classical minimum mean squared error technique (MMSE) an ellipse was  fitted to each structure.

\subsubsection{Automatic search: ``Learning'' phase}

Since the algorithm developed to carry out an automatic search requires  setting  up  different thresholds to run  in an efficient way, a pilot run of this algorithm was made in the same data cubes as are used to construct the visual catalog. 
By iteratively running the algorithm with starting points set at the centers of the available structures in the initial visual catalog, the different thresholds were tuned in order to guarantee a maximum rate of detection and, at the same time, to maximize the similarity between the detected surrounding walls and their associated ellipses defined in the initial visual catalog.

In the following section, STEP 2, the threshold parameters are precisely defined. It is noted that, a fine tuning of these thresholds allowed us to detect a maximum of approximately $80\%$ of those features listed in the visual catalog.

\subsubsection{Automatic search: Blind running phase}

The automatic search for \hi\, supershell candidates  was carried out in the following steps:

\begin{itemize}
\item \textbf{STEP 1: To search for local minima pixels in averaged channel maps.}
Since we need to find structures observed in at least five velocity channel maps, instead of working on single slices at a given velocity, we average five consecutive channel maps. More specifically, at every velocity channel $v_0$, we compute the average by using slices at velocities $v = v_0 + n\delta_v$ with $n=-2,-1,0,+1,+2$ and $\delta_v = 1.031$ \kms\, (channel separation).

In every averaged map we identify those pixels that belong to a local minimum, i.e., a set 
of pixels  whose surrounding pixels have a higher value of brightness temperature.
\\

\item \textbf{STEP 2: To find surrounding walls.}

At each detected local minimum pixel, and for different position angles $\theta$, the algorithm computes the brightness temperature profile T$_b (r,\theta)$ along radial lines having their origin at the corresponding minimum pixel (see Fig. \ref{fig:example}b). A total of 100 radial lines being separated by $\Delta \theta = 3\fdg6$, are analyzed for every local minimum pixel. 
Along each T$_b (r,\theta)$ a maximum (or peak) is present
 if the following conditions are simultaneously fulfilled:

\begin{enumerate}
 
\item Along each  T$_b (r,\theta)$, a ``big slope point'' is found at a point $r = r_{slp}$, if the slope at that point exceeds a predetermined slope threshold (T$_{slp}$) (i.e. if $\frac{d \mathrm{T}_b (r_{slp},\theta)}{dr} > \mathrm{T}_{slp}$, see Figure \ref{fig:example}b). From the ``learning phase'', the threshold slope was set to $\mathrm{T}_{slp}=0.2$ K/px

\item At a certain point $r=r_M$ ($r_M>r_{slp}$), T$_b (r_M,\theta)$ reaches a local maximum. This local maximum is defined in such a way that its brightness temperature T$_b (r_M,\theta)$ exceeds by at least a threshold $\delta_T$ the brightness temperature of both T$_b (r_{M-1},\theta)$  and T$_b (r_{M+1},\theta)$; i.e., T$_b (r_M,\theta) > $  T$_b (r_{M\pm 1},\theta) + \delta_T$. 
This criterion is set to avoid misidentifying of the local maximum due to noise fluctuations (see Fig. \ref{fig:example}b). The value of $\delta_T=0.4$ K was set during the ``learning phase''. It  should be noted that this value is five times higher
than the rms noise of the LAB survey (0.07-0.09 K) \citep{kal05}.

\item To avoid detecting  structures that may be mostly the consequence of the algorithm picking up pixels whose peak temperature is only a few times the  local \textit{confusion} level in brightness temperature (T$_{\rm conf}$), which arises from the normal structuring of the ISM, (e.g., fake structures), the maximum temperature T$_b (r_M,\theta)$  of each brightness temperature radial profile must always exceed the brightness temperature at the local minimum pixel T$_b (0,\theta)$ by a certain threshold (e.g., T$_b(r_M,\theta) - \mathrm{T}_b(0,\theta) > \mathrm{T}_M(l,b,v)$). The value of T$_M(l,b,v)$ depends on the position of the pixel within the \hi\, data cube. 
At low Galactic latitudes   $\mathrm{T}_M(l,b,v)$ should be higher than at higher ones, because the \textit{confusion} levels are higher in the Galactic plane.
Since an analytical form of T$_M(l,b,v)$ is not available, during the ``learning phase'' we have computed the optimal value of $\mathrm{T}_M(l,b,v)$ corresponding to the location of each feature listed in the initial visual catalog by maximizing the similarity between the detected walls and the associated known structure. 
Since the  structures found  in the visual catalog are mostly confined to the central part of the HI data cubes, the obtained  values of  $\mathrm{T}_M(l,b,v)$   cannot be used outside the region defined by the structures belonging to the visual catalog.
Thus, to estimate $\mathrm{T}_M(l,b,v)$  in these locations, new data cubes were created having the same coverage in Galactic longitude  than before but covering 100$^\circ$ in Galactic latitude ($-50^\circ \le b \le 50^\circ$).
Then, for each of these data cubes, six square regions of $2^\circ$ in size were selected and used to compute T$_{\rm conf}$. Since we need to estimate T$_{\rm conf}$ at the boundaries  of the data cubes, the center of each of the six regions are located at Galactic longitudes $l_l + 1^\circ$ and $l_h - 1^\circ$ and at  Galactic latitudes $b_l +1^\circ$, $b_h -1^\circ$, and  $b=0^\circ$, where the subscripts $l$ and $h$ refer to the lowest and highest values (for a given data cube), respectively.
Then,  for three different  velocity channels (two corresponding to the extreme velocities of the data cube and the third one at an intermediate value) within each region, we compute  the standard deviation ($\sigma$) of the brightness temperature  and adopt T$_{\rm conf}$ = $\sigma$ and $\mathrm{T}_M(l,b,v)$ = 3 T$_{\rm conf}$.

Therefore, the corresponding threshold T$_M(l,b,v)$ at any position $(l,b,v)$ is obtained by a 3-D interpolation of all available estimates, i.e., those corresponding to structures available in the visual catalog plus those derived for the boundaries of the data cube.

For every detected structure, it is desirable that the points defining the \hi\, ``wall''  are  ``well behaved''. It means they should follow a more or less ordered pattern (i.e., the distance of a given point of the \hi\, wall to the center of the structure should not drastically differ from the distance of the precedent or posterior). Under this assumption, the algorithm requires that all the points  defining the \hi\, wall should be located at a distance such that $d_n$ (the distance of the n-point) must verify that 

\begin{equation}\label{condition}
\centering
d_{min} < d_n < d_{max},
\end{equation}

\noindent where the parameters $d_{min}$ and $d_{max}$ are defined for every object by performing a principal component analysis (PCA) on the complete set of local maxima defining  the \hi\, wall of a particular structure. Using PCA the algorithm  finds both the direction along which the set of \hi\, maxima points defining the wall achieve their maximum variance, and its orthogonal direction. We denote by $\sigma_1$ and $\sigma_2$  the standard deviations along these directions. By fine-tuning  the relationship between 
$d_{min}$, $d_{max}$, $\sigma_1$ and $\sigma_2$  during the ``learning phase''  of the algorithm, it was found that $d_{min} \sim 0.5\, \sigma_2$ and $d_{max} \sim 2\,  \sigma_1$.  After each cycle all points  that do not  fulfill condition \ref{condition}  are dropped from the original set of points  defining the \hi\, wall. The PCA  iteration  cycle stops when  all remaining  points fulfill  condition \ref{condition}.\\

\end{enumerate}
 
Finally, the algorithm will assume that it has found an \hi\, wall around a brightness temperature minimum when at least 50 individual \hi\, peaks have been identified, which corresponds to half of the total radial directions, even in the case where they do not  belong to consecutive radial lines.\\
 
\begin{figure*}
\center
\includegraphics[width=16cm]{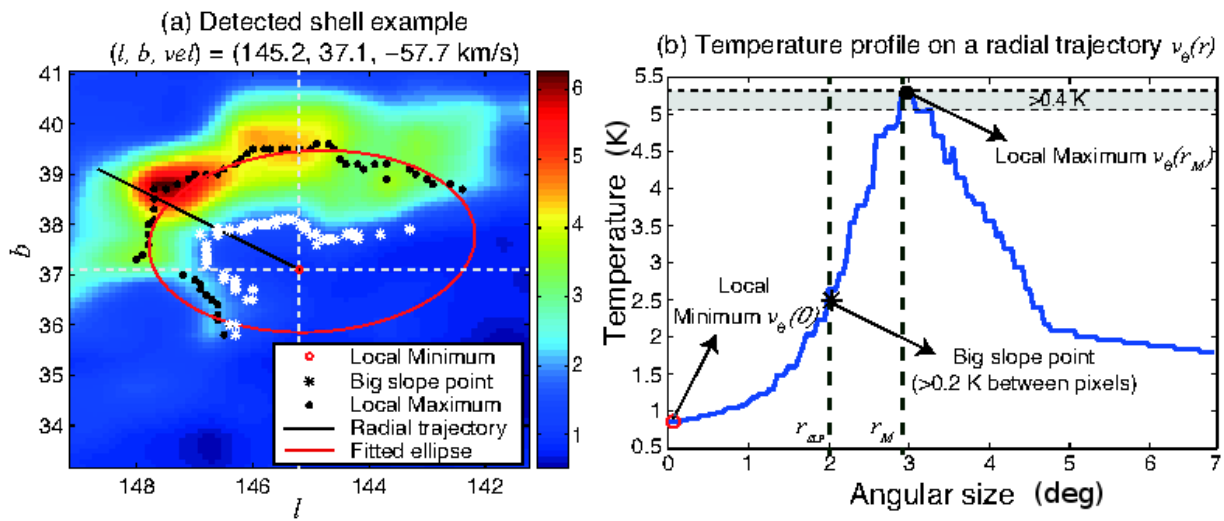}\\
\caption{Example of a shell detected on the average of five contiguous velocity channels (a) centered at the systemic velocity ($-57.7$ \kms), and one corresponding temperature profile (black straight line  in (a) panel) where the big slope and the local maximum points  are identified (b).}
\label{fig:example}
\end{figure*}

\item \textbf{STEP 3: To unify multiple local minima observed at the same velocity interval.}

Sometimes, several local minima are found inside a given supershell candidate. To avoid multiple listings of the ``same'' structure with slightly different central pixels, the algorithm compares the HI ``walls'' associated with every central pixel and keeps the minimum among all the central pixels. To measure  the ``discrepancy'' between the two sets of points, let us say $S_1 =\{\mathbf{x}_1,\mathbf{x}_2,\dots,\mathbf{x}_{N}\}$ and $S_2 =\{\mathbf{y}_1,\mathbf{y}_2,\dots,\mathbf{y}_{M}\}$, where $N$ and $M$ are the number of points in $S_1$ and $S_2$, respectively ($N\le M$); for each point $\mathbf{x}_i$ in $S_1$, we measure the Euclidean distance to its closer point in $S_2$ denoted by $d(\mathbf{x}_i,S_2)$ and  compute the mean discrepancy between $S_1$ and $S_2$ as follows $D = \frac{1}{N}\sqrt{\sum_{i=1}^N d^2(\mathbf{x}_i,S_2)}$. We decide that the two sets of points are assumed to belong to the same structure if $D$ is smaller than a threshold that was empirically set equal to $0.2$. In case of coincidence, only the structure with the minimum central pixel is kept.
\\

\item \textbf{STEP 4: Ellipse fit.}
In this step the algorithm fits an ellipse to the entire set of points defining the \hi\, walls, by using a classical MMSE technique. Every ellipse is characterized by  the following set of parameters: i) the coordinates of the centroid $(l_e, b_e)$; ii) the major (a) and minor (b) semi-axes; and iii) the inclination angle ($\phi$) between  the Galactic plane and the major axis. This angle is measured counterclockwise from the Galactic plane.\\

\item \textbf{STEP 5: Selection of structures according to their sizes.}

Finally,   based on the value of the major axis derived in the previous step and assuming the galactic rotation model of \cite{fic89}, the linear size  of the major axis  is derived. Only those structures having a major axis  larger than  200 pc are retained  in our catalog.\\

\end{itemize}

\subsubsection{Final visual inspection}

To obtain  the final set of structures, after implementing the automatic search, we  made a new  visual inspection (again using  the package \textit{kvis} of KARMA) of all the detected structures.
Considering that a structure will change its shape as one steps through different velocity channels, the structure will be seen in a $(l,b)$ diagram as a set of concentric rings at different radial velocities, in the  ideal case of a symmetric expansion. At the systemic velocity, $vo$, the ring should  attain its maximum dimensions, while  the \hi\, emission distribution should look like a cap at extreme velocities. According to this, we expect  a change in the shape of the structures and   a  continuity in the walls of the observed features at different velocity channels. 
Though  this ideal behavior is usually not observed, in this final visual check we discard those features whose behavior  strongly departs from the one described above.

As an example, we mention structures whose centers in consecutive channel maps ``jump'' from one position to another in a random pattern, or structures whose angular diameters vary in an unrealistic physical way.
At the end of this last stage of the catalog elaboration, we  cataloged a total of 566 supershell candidates, 347 in the second Galactic quadrant and 219 in the third one.

For all these structures, the velocity extent  ($\Delta v$) are determined by visual inspection. 
Under the assumption of a symmetric expansion, the expansion velocity ($v_{exp}$) of a shell is estimated as half of the velocity extent of the shell $v_{exp} = \Delta v/2$. The systemic velocity ($v_0$) of each structure is determined as the central velocity channel over the velocity extent.
The effective radius of the structures are derived as the geometrical mean of the semi-major (a) and semi-minor (b) axes of the fitted ellipse, $R_\mathrm{eff} = \sqrt{\mathrm{a} \times \mathrm{b}}$.\\

Table \ref{table-supershells1}, available at the CDS\footnote{Centre de Donn\'ees astronomiques de Strasbourg. http://cdsweb.u-strasbg.fr/}, contains 
all the parameters of the detected supershell candidates. The information is given as follows:\\
{\it Column 1}: the name of the supershell candidate.  using the code, GS\,$ll \pm bb \pm v_0$, where GS stands for Galactic shell, $ll$  and $bb$ are the Galactic longitude and latitude of the center of the fitted ellipse, respectively, and $v_0$ is the systemic velocity of the structure.\\
{\it Columns 2 and 3}: the Galactic longitude and latitude of the center of the fitted ellipse, in degrees. The uncertainty in the values estimated for the centroid of the ellipses is about $\pm 0.2^\circ$.\\
{\it Column 4}: the systemic velocity ($v_0$) in units of  \kms. The uncertainty of  this velocity is equal to the velocity resolution of the data ($\pm 1.3$ \kms).\\
{\it Column 5}: the heliocentric distance, in kpc. The mean weight distance  error for all the sample is about $21 \%$.\\
{\it Columns 6 and 7}: the Galactic longitude and Galactic latitude (in degrees) of the local minimum.\\
{\it Columns 8 and 9}: the major and minor semi-axes, in degrees. The uncertainties in the semi-axes are about $10 \%$.\\
{\it Column 10}: the inclination angle ($\phi$)  (major axis inclination relative to the  Galactic plane, this angle is measured counterclockwise form the Galactic plane). See Section \ref{phi-distribution} for uncertainties in $\phi$.\\
{\it Column 11}:  the effective radius ($R_\mathrm{eff}$) of the structures, in pc. The uncertainties in these values come from the uncertainties in the distances and  angular semi-axes, and are around 16 \%.\\
{\it Column 12}:  the velocity extension ($\Delta\, v$), in \kms, of the structures. The uncertainty is about $2.6 $ \kms.\\
{\it Column 13}: the total number of full quadrants: i.e.,  each structure has been divided, from the center of the fitted ellipse, into four quadrants. We counted the number of quadrants that present \hi\, emission related to the wall of the structure (full quadrants). See Section \ref{morphology}.\\
            
An excerpt from the table is shown in Table \ref{table-supershells1}.

\begin{table*}
\center
\caption{Excerpt form the table of the supershell candidates available at the CDS$^2$. The asterisk after the supershell name indicates that the structure is a chimney candidate (see section \ref{morphology}).}

\label{table-supershells1}
\begin{tabular}{lrrrrrrrrrrrr}

\hline
\hline
Supershell&$l_e$ &$b_e$ & $v_0$& $d$ &$l_0$ & $b_0$ & a & b & $\phi$& $R_{ef}$ & $\Delta v$ & N \\
candidate &($^\circ$) &($^\circ$) & \kms & kpc &($^\circ$)& ($^\circ$) & ($^\circ$) & ($^\circ$) &($^\circ$) & pc & \kms &  \\
\hline

\ GS\,089$-$21$-025$*&88.9&-21.0&-24.7&4.3&91.1&-21.4&6.9&4.0&-1.5&394&11.3&3 \\
\ GS\,090+09$-077$&90.3&9.0&-77.3&9.9&90.6&10.3&3.5&1.9&87.1&449&21.6&4\\
\ GS\,091$-$01$-056$&90.6&-0.8&-55.7&7.4&90.2&-1.3&1.6&1.3&-6.7&183&10.3&4\\
\ GS\,091+06$-115$&90.7&5.8&-115.4&15.6&90.7&6.3&2.3&1.6&-36.1&521&9.3&3\\
\ GS\,091$-$04$-067$*&91.0&-3.6&-67.0&8.4&91.9&-4.0&4.6&2.7&1.7&512&35.0&3\\
\hline
\end{tabular}
\end{table*}

\section{Selection effects}\label{selection}

Due to the location of the Sun (our observing place) within the Milky Way, our  catalog suffers from distance dependence.
Our criteria for identifying candidate features  introduce selection effects  in the calatog that must be taken into account  when attempting  to derive their large scale properties.

\begin{figure}
\vspace{1cm}
\center
\includegraphics[angle=0,width=9cm]{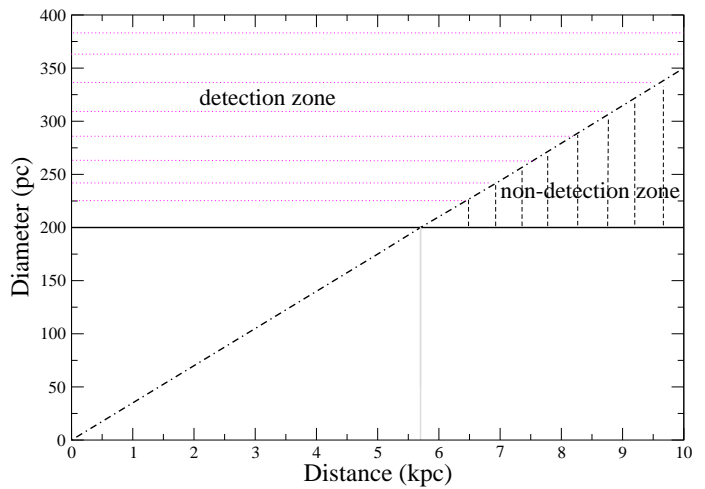}\\
\caption{Scheme showing selection  effects. The black dashed dotted line shows $2^\circ$ angular size structures at different distances from the Sun. The full horizontal line represents structures with linear sizes of 200 pc. The gray line at 5.7 kpc from the Sun marks the limit of detection of a $2^\circ$ structure. The area marked with black dashed lines shows the non-detection region. The area marked with  dotted lines shows the detection zone.}
\label{nondetectionZone}
\end{figure}

As a consequence of the selection criterion c)  (the structure must have a  minimum angular diameter of $2^\circ$), there will be structures that, even  when having  diameters greater than  200 pc, will not be detected by the algorithm if at their corresponding distances  the  angular diameters  are lower than $2^\circ$. For instance, a structure with a diameter of 200 pc located at 5.7 kpc from the Sun will be in the limit of detection, with an angular size of $2^\circ$.
In Fig. \ref{nondetectionZone} the region  that suffers for this effect is labeled as ``non-detection zone''. As can be inferred, beyond 5.7 kpc from the Sun, the catalog of   \hi\, supershell candidates will not be complete.

 On the one hand, large linear size features located close to the Sun will have  large angular diameters and will be missed by the detection algorithm because they are too big  compared to the field size of the data cube where the algorithm is run. In other words, structures  with  linear sizes larger than  1 kpc that are  closer to the Sun than $1.1$ kpc  will have  angular dimensions greater than $\sim 50^\circ$ and  will not be detected by our algorithm. 
 Besides, at 2 kpc away from the Sun, a structure having the minimum required size (200 pc) to be cataloged as  a supershell candidate will have a  large angular diameter of  $ 5.7^\circ$, raising the possibility of having in  its interior a lot of small scale structures.
As a consequence, our algorithm may fail to identify large angular structures.

\section{General properties of the supershell candidates}

As mentioned before, after following all the steps described in Section 3, we have detected a total of
347 structures in the second Galactic quadrant ($90^\circ < l <  165^\circ$) and 219 in the third one ($195^\circ < l <  270^\circ$). In the following we analyze each quadrant separately.

\subsection{Supershells morphology}\label{morphology}

One characteristic that makes our algorithm  different from other automatic implementations \citep[e.g.,][]{ehl05,ehl13}  is that it is able to detect structures that are not completely surrounded by walls of \hi\, emission (``closed''). To decide whether a supershell candidate is completely  ``closed'' or not (``open''), we divided each supershell candidate into four quadrants centered on the center of the fitted ellipse and determine how many quadrants show \hi\, emission (local maxima).  To tag a given quadrant as ``full'' or ``empty'', our algorithm counts how many local maxima points ($npts$) are present in each quadrant.  We named the maximum number of points among the four quadrants as $npts_m$.
Afterwards, for each structure,  the quadrant having the highest number of  local maxima points is selected.
Then, a quadrant is considered ``full''  if the number of local maxima ($npts$) in that quadrant is greater than $npts_m - 3 \sigma$, where $\sigma = \sqrt{npts_m}$. Otherwise  the quadrant  will be considered ``empty''.

A careful inspection of all the structures (347 in total) found in the second Galactic quadrant reveals that  182 ( $\sim 52$ \% of the total) are completely 
closed structures, 120 have one  quadrant empty, 32 present two  quadrants empty, and 13 have all quadrants empty  but one. 
As for the third Galactic quadrant, 126 ($\sim 58$ \% of the total) out of the 219 detected structures are completely closed structures, 62 have one  quadrant empty, 24 have two of the quadrants empty, and only 7 have one of the quadrants full.

Open structures may  play an important role in the Galaxy because it is believed that these structures, when open towards the Galactic halo,  may play a role in injecting material from the Galactic plane into the Galactic halo. These structures, open towards high Galactic latitudes may represent candidates features to be identified as ``galactic chimneys''.
To look for these features among all the detected supershell candidates, we selected those having one  quadrant ``empty'' and analyzed whether it is  open toward the halo or toward the Galactic plane.

To this end, we have only taken those structures into account that have a galactocentric distance lower than  15 kpc. This restriction is imposed because beyond  15 kpc, the warp  of the Galactic plane becomes important and complicates the identification of the structures whose opening is mostly directed away from the Galactic plane defined in the classical way ($b = 0^\circ$). Structures with two or three ``empty'' quadrants are not taken into account because they only represent  a very small fraction  ($\sim 13-14$\%) of the overall number of the structures present in the catalog.

A total of 120 structures were found to have one quadrant ``empty'' and a  kinematic galactocentric distance smaller than 15 kpc. Among them about 67 \% are open towards the Galactic halo, and may be classified  as ``galactic chimney'' candidates. They are identified in the online table with an asterisk after the supershell's name. As an example see the  excerpt of the online table (Table \ref{table-supershells1}).
On the other hand,  we have found that 21\% of all detected structures having galactocentric distances lower than 15 kpc are classified as ``galactic chimney'' candidates.

\subsection{Distribution in the sky}

\begin{figure*}
\center
\includegraphics[width=15cm]{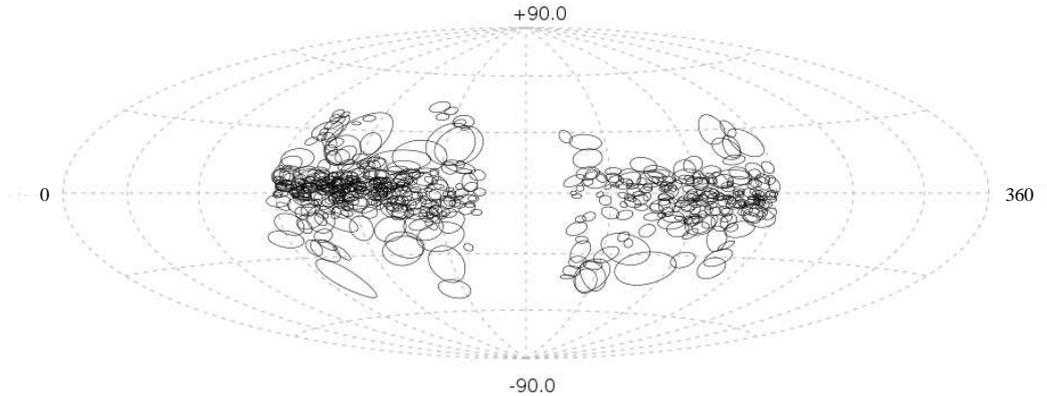}\\
\caption{Supershell candidates in a Hammer-Aitoff projection, centered on ($l, b$) = ($180^\circ, 0^\circ$).}
\label{hammer}
\end{figure*}

\begin{figure}
\vspace{0.7cm}
\center
\includegraphics[width=8cm]{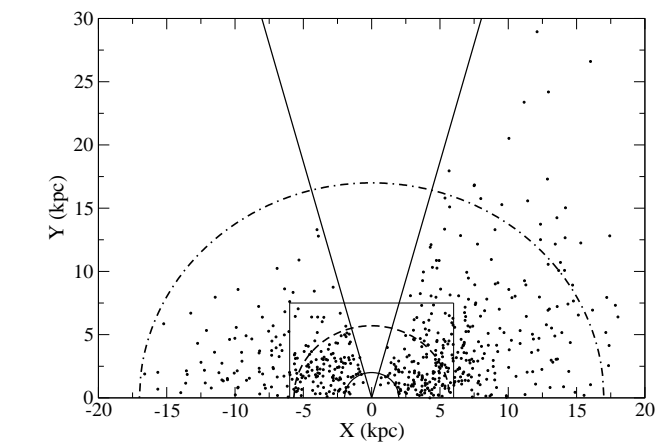}\\
\vspace{1.1cm}
\includegraphics[width=8cm]{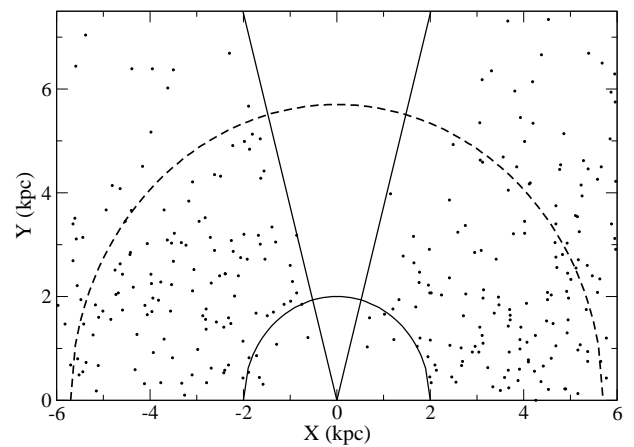}\\
\caption{Distribution of the supershell candidates in a polar diagram. Dots correspond to all the structures detected. \textit{Top panel}: the 2 kpc ring around the Sun is marked by a continuous black line semi-circumference.  The 5.7 kpc and 17 kpc rings  are marked by dashed and dashed-dot semi-circumferences, respectively. The $\pm 15^\circ$ region around the anti-center is delimited by straight  black lines. \textit{Bottom panel}: zoom of the area delimited by the rectangle in the top panel.}
\label{coordpolares}
\end{figure}

The fitted ellipses of the supershell candidates  are plotted in Fig. \ref{hammer} in a Hammer-Aitoff projection. As expected, most of the structures are located near the Galactic plane.
Figure \ref{coordpolares} shows all the centroids of the cataloged features in a polar diagram. The Sun is located at  position (X, Y) = (0, 0). Clearly enough, within 1.5--2 kpc from the Sun, the number of cataloged structures is much smaller than beyond this distance (see Fig. \ref{coordpolares} bottom panel).
Very likely this  reflects a limitation (or selection effect) of our identification algorithm (see Section \ref{selection}).

Towards the anti-center ($l = 180^\circ$), the galactic rotation models predict a very small gradient of the radial velocity with distance. Furthermore,  non-circular motions may be comparable to the supershell radial velocities.
Since both facts will make the kinematic distances of the supershell candidates quite uncertain, those structures whose centroids fall between $165^\circ \leq l \leq 195^\circ$ are not listed in our catalog. 
The  straight lines  having their origin at (X, Y) = (0, 0) mark these boundaries. 

 Figure \ref{coordpolares} also shows that supershell candidates are identified at large distances ($\sim 32$ kpc) from the Sun in the second Galactic quadrant, while in the third quadrant there are not supershell candidates located beyond 17 kpc from the Sun. This finding agrees with \citet{ehl13}. The explanation of this effect is far from clear. This striking difference between both quadrants is unexpected if both quadrants were similar from a geometrical point of view. Certainly, in regards to the \hi\, supershells, a more thorough study of the external part of the Milky Way is needed.

\subsection{Statistics}

In the following, to derive the statistical properties of the supershell candidates  we  only use a subset of the detected structures to minimize the selection effects present in the catalog. We have only considered structures that fulfill the following  conditions: $i$) that are located at distances smaller than 5.7 kpc from the Sun (see Section \ref{selection}), and $ii$) that are completely closed or  have  only one  quadrant ``empty''. 
A total of 190 structures out of the 566 cataloged meet these criteria,  93 in the second Galactic quadrant and 97 in the third one.

Even though we have mentioned before that most of the structures having distances less than 2 kpc will be missed by our detection algorithm. Bearing in mind that the number of these structures represents only 5\% of the cataloged structures having $d < 5.7$ kpc, we have included them in all our statistical analysis, except those related  to the galactocentric distribution analysis (see Section \ref{gd}).

\subsubsection{Size distribution}

\begin{figure}
\center
\includegraphics[angle=0,width=8cm]{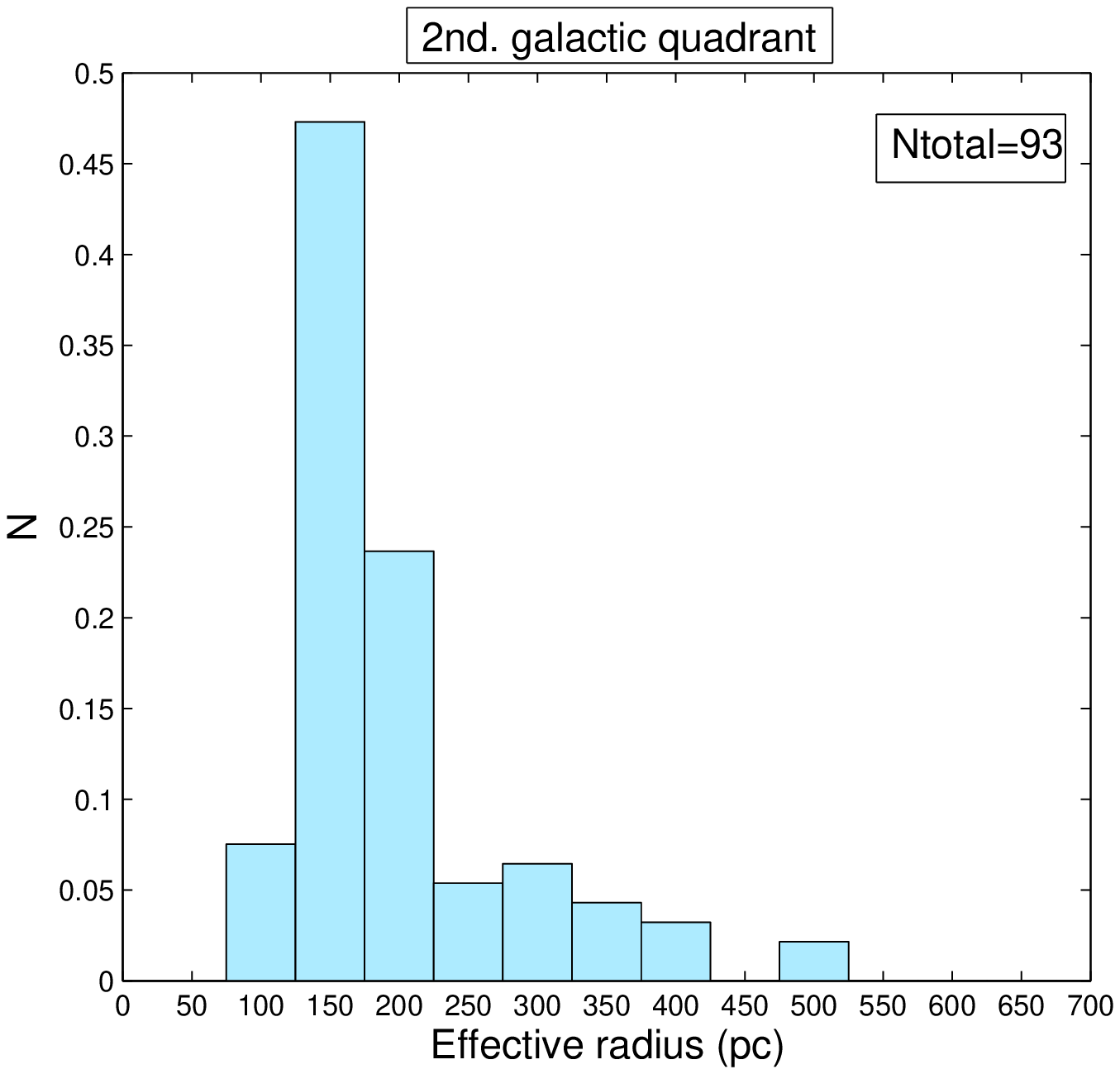}\\
\includegraphics[angle=0,width=8cm]{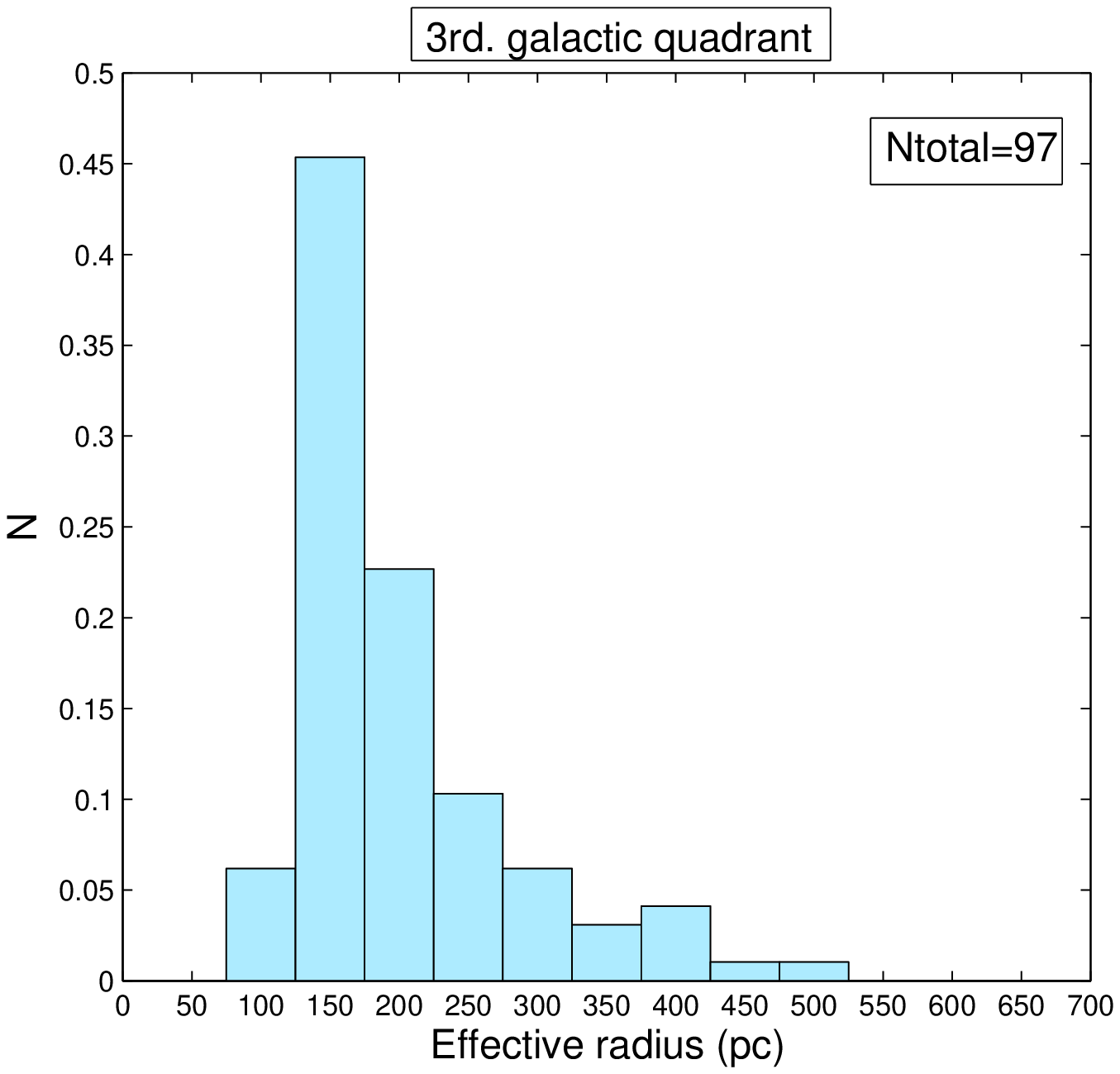}\\
\caption{Distribution of the effective radius of the supershell candidates in the 2nd Galactic quadrant (\textit{upper panel}) and in the 3rd Galactic quadrant (\textit{lower panel}).}
\label{reff}
\end{figure}

As mentioned above,  kinematic distances ($d$) and physical sizes of the supershell candidates are determined after  using the  rotation curve of \citet{fic89}. In Fig. \ref{reff} (upper and lower panels) the $R_\mathrm{eff}$ distribution is shown for the second and third Galactic quadrants, respectively.  The weighted  mean value of $R_\mathrm{eff}$ in the second Galactic quadrant is  $158 $ pc with a dispersion  of $47$ pc, and is  $ 163$ pc with a dispersion of $49$ pc for the third one. The lack of structures with effective radius less than  100 pc is a consequence of our selection criteria.

\subsubsection{Distribution of the position angles ($\phi$)}\label{phi-distribution}

\begin{figure}
\center
\includegraphics[angle=0,width=8cm]{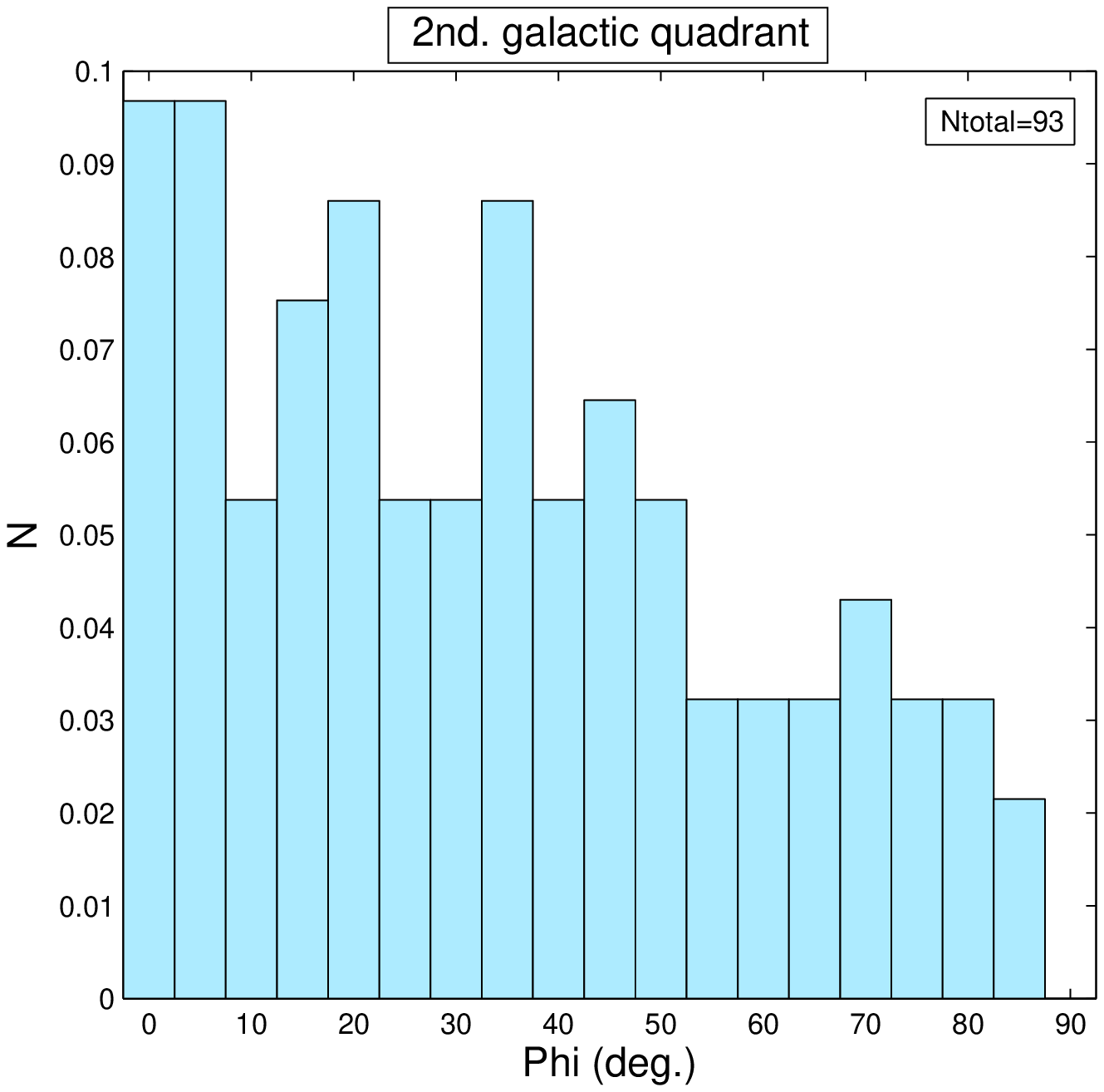}\\
\includegraphics[angle=0,width=8cm]{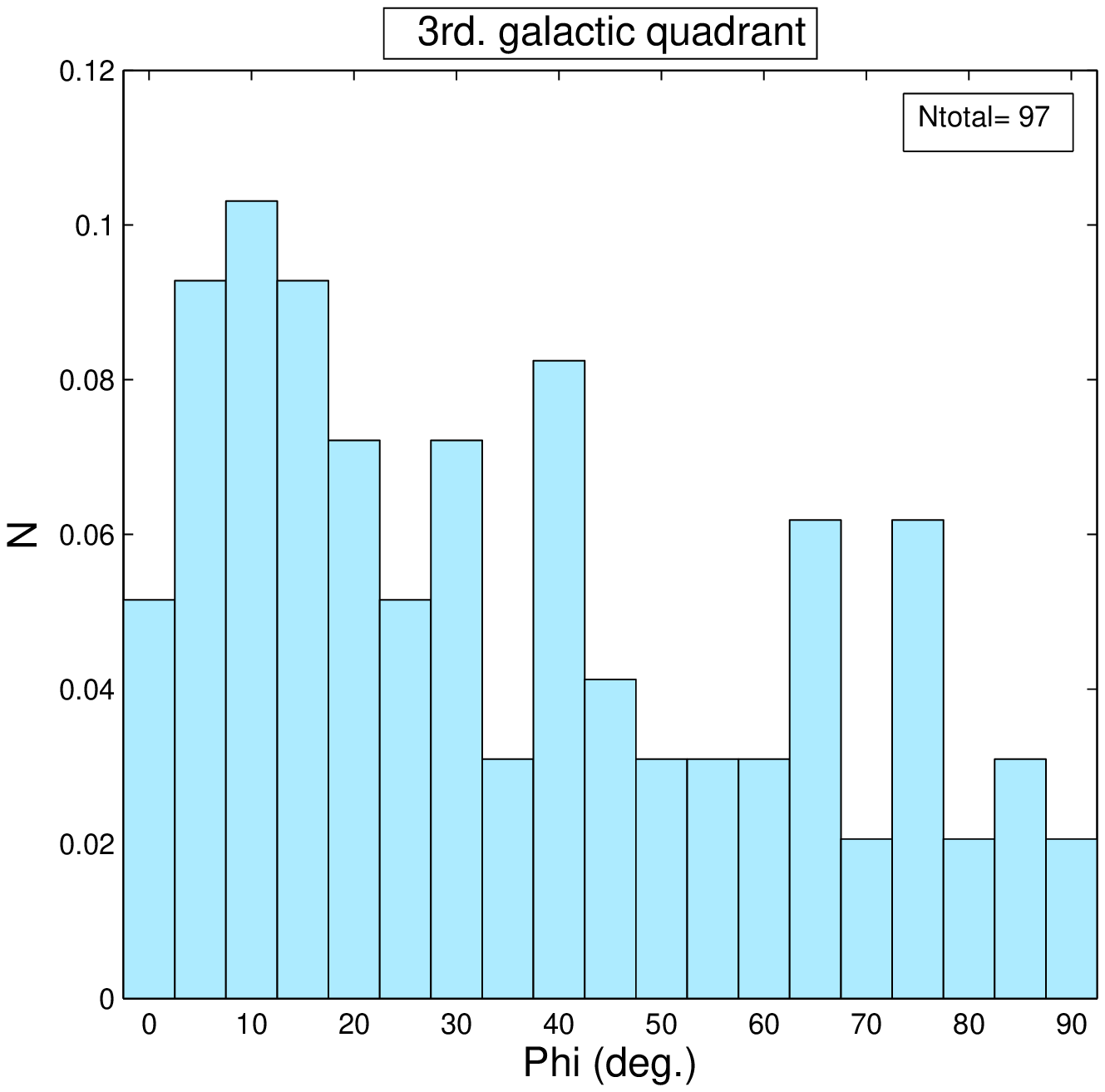}\\
\caption{Relative galactic distribution of the inclination angles ($\phi$) between the major axes of the \hi\, supershell candidates and the Galactic plane for the 2nd Galactic quadrant (\textit{upper panel}) and  the 3rd Galactic quadrant (\textit{lower panel}).}
\label{phi}
\end{figure}

Figure \ref{phi} shows the distribution of the inclination angles $\phi$ of the supershell major axes relative to the Galactic plane.  The error in the value of $\phi$  has  a strong dependence on $\phi$ and a minor dependence on the major semi-axis $a$, according to the following relation:

$$ \sigma_\phi^2 = \frac{1}{(a\, sin(\phi))^2} \left[ \left(\frac{1}{|1-2\,sin(\phi)^2|} + cos(\phi)^2 \right)\, \sigma_a^2 + \sigma_{l_0}^2 \right],  $$

\noindent where $l_0$ is the Galactic longitude of the center of the fitted ellipse and  $\sigma_{l_0} = 0.2$. The error in $\phi$ diverges for $\phi = 0^\circ$ and for $\phi= 45^\circ$.

Given  that the \hi\, density varies  with the Galactic plane height ($z$) such that the \hi\, density decreases as $z$ increases \citep[they found a  scale height of 403 pc]{dic90}, we would expect that structures expand more rapidly in the direction perpendicular to the Galactic plane than along it. In this context, in case the Galactic density gradient was playing a major role in determining the overall shape of the expanding structures, one would expect  the major axis to be mostly perpendicular to the Galactic plane. To analyze this hypothesis, we  study the distribution of the angles $\phi$. 
We consider that the major axis is parallel to the Galactic plane if $\phi < 45^\circ$ and otherwise if $\phi > 45^\circ$. Bearing in mind the errors involved in determining  $\phi$, and taking a 5 $\sigma_\phi$ confidence level into account, those structures having  $42\fdg5 \leq \phi \leq 47\fdg5 $ were not  considered  (these structures are the ones located in the bin centered at $45^\circ$ in Fig. \ref{phi}),  because in these cases a structure considered parallel could also be perpendicular taking into account the  involved errors.

In this way, roughly $ 70\%$ of the structures have their major axes parallel to the Galactic plane for both quadrants.
This result agrees with the one found by \cite{ehl05,ehl13} where they state that the majority of the detected structures are elongated in the longitude direction rather than in latitude.

\begin{figure}
\center
\includegraphics[angle=0,width=8cm]{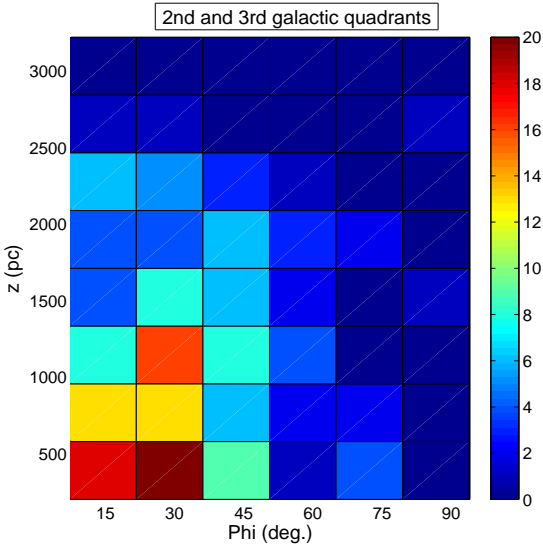}\\
\caption{2-D histogram with $\phi$ in degrees and the Galactic plane height ($z$) in pc. The color bar indicates the number of structures in each bin.}
\label{hist2d}
\end{figure}

To detect a difference in the orientation of the major axis according to the location of the structures with respect to $z$, we made a 2-D histogram (see Fig. \ref{hist2d}). There, it is  shown  that most of the structures have their major axes  oriented parallel to the Galactic plane, independently of the value of $z$, i.e. structures with the major axis oriented parallel to the Galactic plane are distributed in a wide range of $z$. From the 2-D histogram we can conclude that there is not a significant  change in $\phi$ when $z$ increases.

It is important to mention that the magnetic fields could play an important role in the evolution of the supershells. Taking magnetic fields and the density stratification in the Galactic disk into account, a 3-D numerical magnetohydrodynamical simulation has been done by \cite{tom98}  where the author concludes that a magnetic field running parallel to the Galactic disk has the effect of preventing the structure from expanding in the direction perpendicular to the field when the magnetic field  has a larger scale height than the density. On the other hand, \cite{tom98} has developed  a model where  the magnetic field strength decreases in the halo as the square root of the density, and found that in this case the structure may eventually blow out.

Based on these findings, the dynamical evolution of the supershell candidates seems to be  less affected by the \hi\, density gradient  present in the direction perpendicular to the Galactic plane than by the galactic magnetic fields running parallel to the Galactic disk.

\subsubsection{Eccentricity of the supershell candidates}

\begin{figure}
\center
\includegraphics[angle=0,width=8cm]{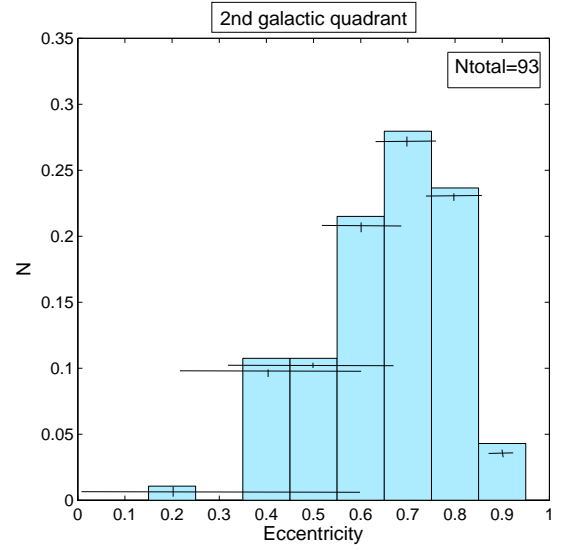}\\
\includegraphics[angle=0,width=8cm]{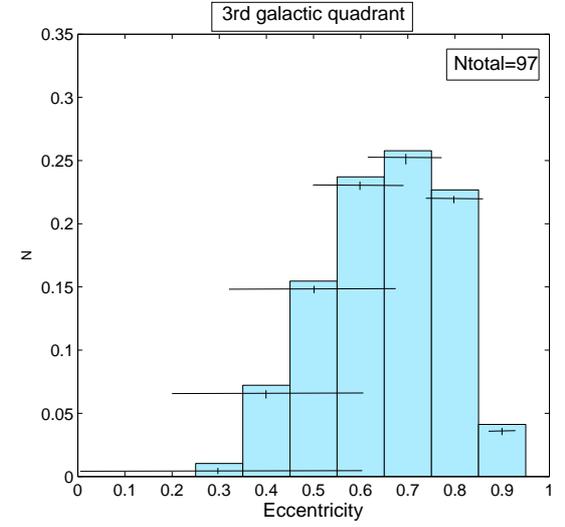}\\
\caption{Distribution of the eccentricities of the supershell candidates in the 2nd Galactic quadrant (\textit{upper panel}) and in the 3rd Galactic quadrant (\textit{lower panel}). Error bars are plotted for each bin, and a parametric form of the error shows that it is about $\sigma_e = 0.6$ for $e=0.1$ and $\sigma_e = 0.02$ for $e=0.9$.}
\label{excentricity}
\end{figure}

The eccentricity of the structures was calculated from

$$ e = \frac{\sqrt{\rm a^2\,-\,\rm b^2}}{\rm a}$$

\noindent where a and b are the major and minor semi-axes, respectively, of the fitted ellipse. 
 Figure \ref{excentricity} shows the distributions of the eccentricities for the second and third Galactic quadrants.
The error associated with the eccentricity varies for different values of $e$. A mean  error bar corresponding to different values of $e$ is plotted in  Fig. \ref{excentricity}. 
It can be seen that about $98 \, \%$ of the structures are elliptical with   mean weighted eccentricities of  $0.8 \pm 0.1$ for both quadrants.

\subsubsection{Distribution of the structures relative to $z$}

\begin{figure}
\center
\includegraphics[angle=0,width=8cm]{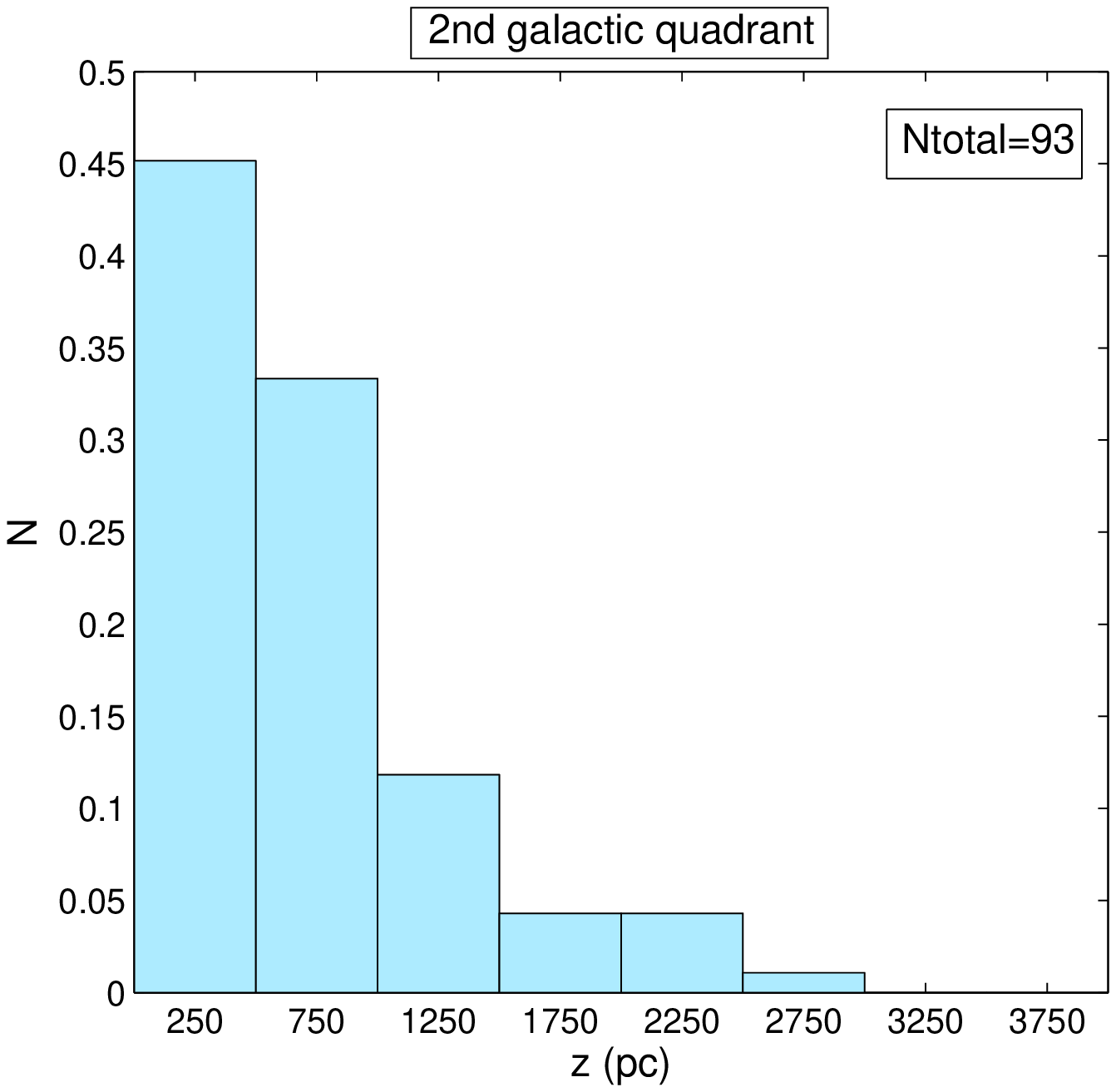}\\
\includegraphics[angle=0,width=8cm]{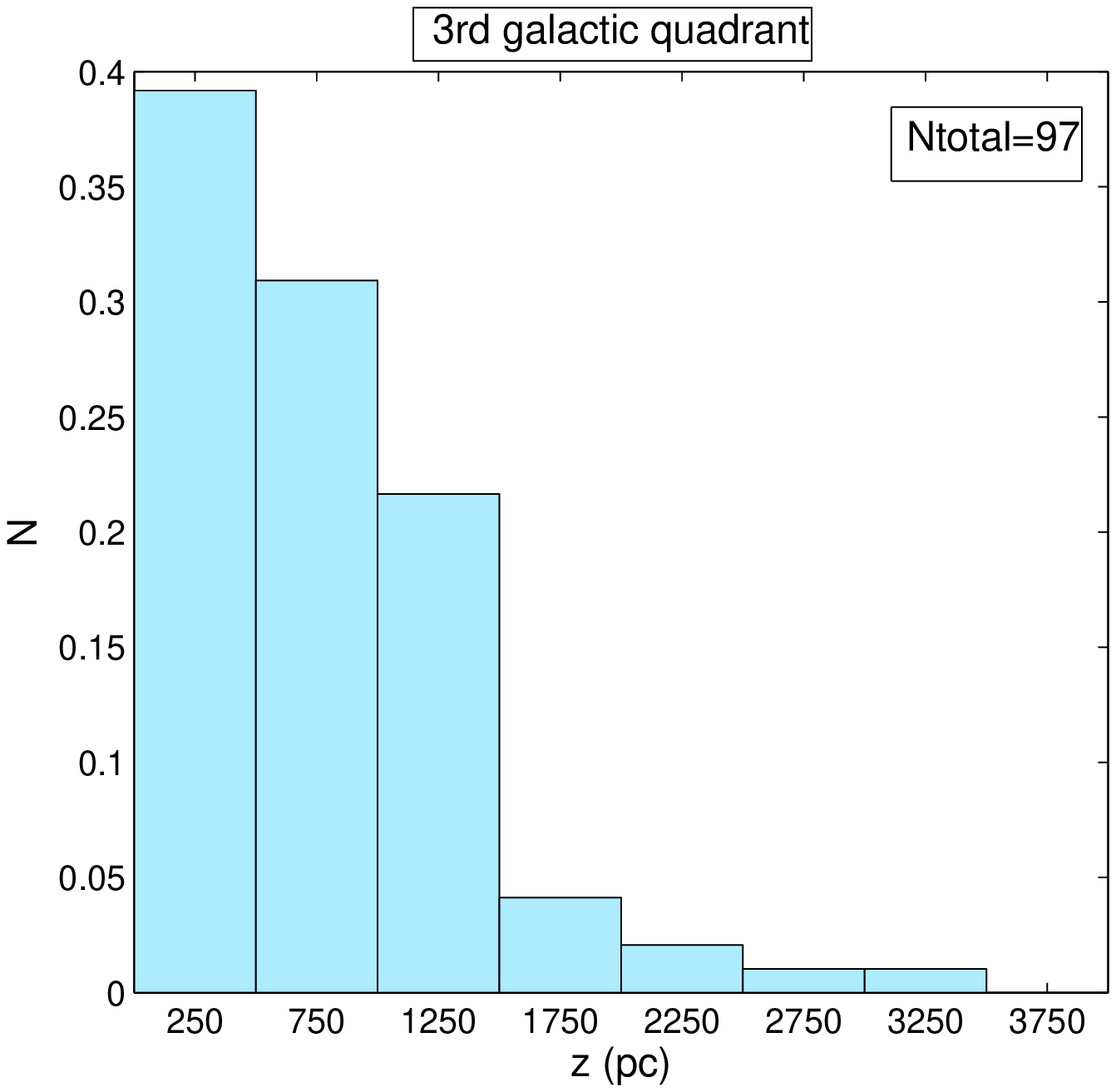}\\
\caption{Distribution of the centroids of the structures relative to $z$ in the 2nd Galactic quadrant (\textit{upper panel}) and in the 3rd Galactic quadrant (\textit{lower panel}).}
\label{z}
\end{figure}

The distribution of the  centroids of the structures with respect to $z$ is shown in Fig. \ref{z}.  Taking the location of the centroids into account, we calculated that  42\% and 39\%  of the structures from the second and third Galactic quadrants, respectively, are confined to  $z \leq 500$ pc. This result agrees, within the uncertainties, with the result obtained by \cite{ehl05}, in the sense that these authors found  that half of all their shells lie in a 1-kpc thick layer.

\subsubsection{Galactocentric distribution}\label{gd}

To attempt  to derive  the distribution of the supershell candidates with respect to the galactocentric distance, the number of structures identified in our catalog has to be corrected for at least two limitations, namely, \textit{a)} the $\pm 15^\circ$ zone of avoidance around  $l = 180^\circ$, and \textit{b)} the failure of our  algorithm to detect structures closer than 2 kpc to the Sun.

\begin{figure}
\center
\includegraphics[width=8cm]{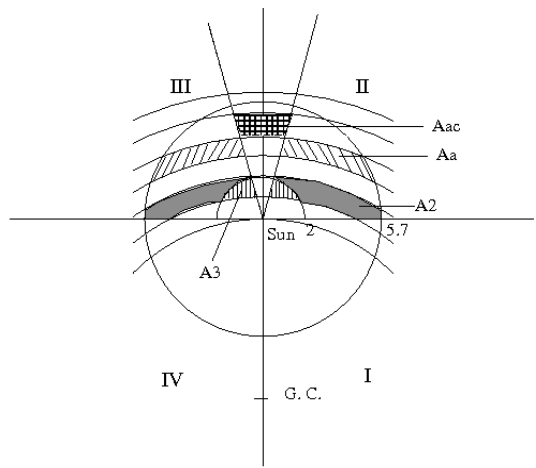}\\
\caption{Scheme of some of the areas $A_a$ (between rings of radius R$_i= 11.5$ kpc and R$_e = 12.5$ kpc, in diagonal black lines), $A_{ac}$ (between rings of radius R$_i= 12.5$ kpc and  R$_e = 13.5$ kpc, in gridded), $A_2$ (between rings of radius R$_i= 9.5$ kpc and  R$_e = 10.5$ kpc, in gray), and $A_3$ (between rings of radius R$_i= 9.5$ and R$_e= 10.5$ kpc, in straight black lines). G.C. marks the position of the Galactic center.}
\label{Aa-Aac-A2}
\end{figure}

Bearing the above set of restrictions in mind, the number of candidate features observed to be  \hi\, supershells must be multiplied by four different ``incompleteness factors'' before attempting  to derive their large scale properties (on galactic scale). To derive the different correction factors, we  divide the Galaxy into a set of concentric rings centered on the Galactic center. Each ring will have an inner radius  of R$_i$ kpc and an outer radius  R$_e$, where R$_e =$ R$_i +1$ kpc. Since we are looking for structures  in the outer part of the Galaxy, the first ring we have to deal with has R$_i = 8.5$ kpc and R$_e = 9.5$ kpc. For a given ring, we define the area between $165^\circ \leq l \leq 195^\circ$ as $A_{ac}$ (see Fig. \ref{Aa-Aac-A2}). Furthermore, we let the area  for those rings having R$_i \geq 10.5$ kpc, outside $165^\circ \leq l \leq 195^\circ$  but within the 5.7 kpc solar semi-circle, be referred to as $A_a$. For those rings having R$_i$ equal to 8.5 kpc and 9.5 kpc, the area comprised between the heliocentric semi-circles of radius 2 and 5.7 kpc and the straight line joining $l= 90^\circ$ and $l = 270^\circ$ will be denoted $A_2$  (see again Fig. \ref{Aa-Aac-A2}), and the area delimited by the zone of ``avoidance'' around $l = 180^\circ$ and the heliocentric semi-circle  of radius 2 kpc will be referred to as $A_3$. The values of the mentioned areas are listed in Table \ref{tabla-areas}.

Considering the above definitions, and assuming that the supershell candidates density (number of supershell candidates per unit area)  is constant within a given galactocentric ring, we defined the correction factors $F_a$ as those to be applied in order to attempt to correct the observed structures for those that should fall within $\pm 15^\circ$ zone of avoidance around $l = 180^\circ$. We also defined the correction factors $F_b$ as those to be applied in order to attempt to correct for those features  that should be  located between the straight lines defining the boundaries of the mentioned zone of avoidance and the semicircle with radius of 2 kpc centered on the solar location. The correction factors are the following: 

$$ F_a = \frac{A_{ac}+A_2+A_3}{A_2+A_3} \qquad \rm{if} \, \, \rm{R}_i   \leq  9.5 \, \rm{kpc}$$

$$ F_a = \frac{A_{ac}+A_a}{A_a} \qquad \rm{if} \, \, \rm{R}_i   \geq  10.5 \, \rm{kpc}$$

$$ F_b = \frac{A_{2}+A_3}{A_2} \qquad \rm{if} \, \, \rm{R}_i   \leq  9.5 \, \rm{kpc}$$

$$ F_b = 1 \qquad \rm{if} \, \, \rm{R}_i   \geq  10.5 \, \rm{kpc}.$$

To derive  the correct number  of \hi\, supershell candidates in the Galaxy for a given galactocentric ring ($N_c$,  given in the ninth column of Table \ref{tabla-areas}), the observed  number of such features ($N_0$, given in the  seventh column of Table \ref{tabla-areas}) must therefore be multiplied by a correction factor,

$$N_c = F\,N_0,$$

\noindent where $F = F_a\, F_b$.

\begin{figure}
\vspace{0.5cm}
\center
\includegraphics[angle=-90,width=8cm]{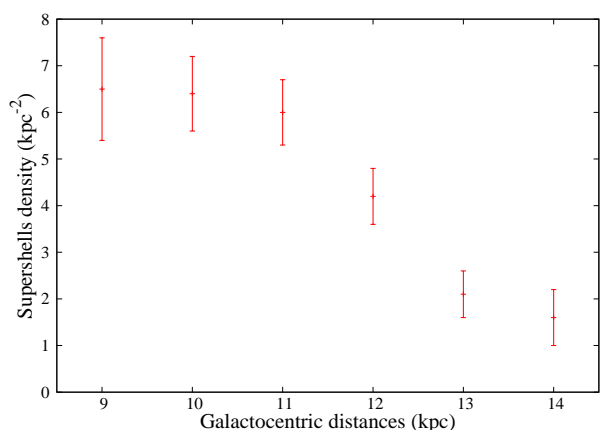}\\
\caption{Surface supershell density  as a function of galactocentric
 distance.}
\label{density}
\end{figure}

Finally, the surface density  of \hi\, supershell candidates for a given galactocentric ring is defined  as the ratio between the corrected number of supershell candidates ($N_c$) and the area of the ring in the outer part of the Galaxy. The corresponding values are listed in Table \ref{tabla-areas}. Figure \ref{density} shows the \hi\, supershell candidates surface density  as a function of the galactocentric distance.

\begin{table*}
\center
\caption{Parameters used to estimate the density of supershell candidates  in the areas $A_{a}+\rm A_{ac}$.}
\vspace{0.3cm}
\begin{tabular}{c c c c c c c  c c c  c c}
\hline
$R_i$ & $R_e$ & A$_{ac}$ & A$_a$ &A$_2$ & $A_3$&$N_0$& $F$ & $N_c$ & Density\\
 (kpc)&(kpc) & (kpc$^2$) &  (kpc$^2$)&(kpc$^2$)&(kpc$^2$) && & &(kpc$^{-2}$)\\
\hline
\hline
 $8.5$& $9.5$ &$ 0.26$ & $5.33$ &   $2.17$&  3.16 &14\tablefootmark{*}&$2.578$&36&  6.5 $\pm$ 1.1\\
 $9.5$ &$10.5$ &$ 0.78 $&$10.13$ & $7.99$ &  2.14 &51\tablefootmark{*}& $1.365$ &70& 6.4 $\pm$ 0.8\\
 $10.5$& $11.5 $& $1.31$& $10.44$ &$-$&  $-$  & 63 &$ 1.125 $& $71$& $6.0 \pm 0.7$\\
 $11.5$& $12.5$ & $1.83 $& $8.9$ &$-$&  $-$  &37&$1.206$& 45& $4.2 \pm 0.6$\\
 $12.5$& $13.5 $&$ 2.36$ &$ 6.21$& $-$&  $-$ &13 &$1.380$ &18&$2.1\pm 0.5$\\
 $13.5$& $14.5$ & $2.02$ &$ 1.82 $& $-$& $-$  &3 &  $2.109$ & 6&$1.6 \pm 0.6$\\
\hline
 \end{tabular}
 \tablefoot{\tablefoottext{*}{For these estimates the structures found within the semicircle of 2 kpc radius around the Sun were not considered.}
 }
 \label{tabla-areas}
 \end{table*}

 Following \cite{ehl13}, but  assuming that the supershell surface density, 
$\Sigma$(R), follows a Gaussian distribution with galactocentric distance R
 given by

\begin{equation}\label{sigma}
\Sigma (R) = \Sigma_0 \times e^{-[(R-R_0)/\sigma_{GS}]^2} \qquad [\mathrm {kpc^{-2}}],
\end{equation}

\noindent where $\Sigma_0$ is the supershell density in the solar neighborhood, $\sigma_{GS}$ is the galactocentric scale length, and {\rm R} and {\rm R$_0$} are the
supershell's and Sun's galactocentric distances, respectively. Fitting Eq. \ref{sigma}
to the surface density given in Table  \ref{tabla-areas} (see last column), we
 derive $\Sigma_0$ = 7.4 $\pm$ 0.5  kpc$^{-2}$ and $\sigma_{GS}$ = 4.4 $\pm$ 0.3 
kpc. The galactocentric radial  scale length is comparable to the one found in previous works
\citep[see][and references therein]{ehl05}, but the estimated surface density of
H{\sc i} supershells in the solar neighborhood is a factor $\sim$ 2 greater than previous estimates \citep{ehl13}. This higher surface density may well be explained by the ability of our procedure  to also identify  structures that are not completely closed. In this context, we would like to point out that the total number of supershells that are either completely closed or have one quadrant ``empty'' (see Sect. 5.1) is almost a factor of 2 larger than the total number of closed features.

\subsubsection{Filling factors}

The filling factors of the supershell candidates are defined as the area ($f_{2d}$) or volume ($f_{3d}$) occupied by the supershells in a given area or volume, respectively.
The area considered was calculated in a semicircle of 5.7 kpc from the Sun, except for the area of $\pm 15^\circ$ around $l = 180^\circ$, and the volume was calculated considering the same area and a Galactic plane height of $\pm 0.5$ kpc.

To estimate $f_{2d}$ filling factor, the structures should be  projected in the mentioned area. The estimated filling factors in the outer part of the Galaxy are $f_{2d} = 0.5 \pm 0.1$ and $f_{3d} = 0.04^{+0.01}_{-0.02}$. These values, within the errors, agree with  those derived by \cite{ehl05}, $f_{2d} = 0.4$ and  $f_{3d} = 0.05$.
Since the filling factors are dominated by larger shells, and bearing in mind that  our method is not able to detect them when located close to the Sun (see Section \ref{selection}), the quoted filling factors may be regarded as lower limits.

\subsubsection{Effective radius versus expansion velocity}

\begin{figure}
\vspace{1.5cm}
\center
\includegraphics[width=8cm]{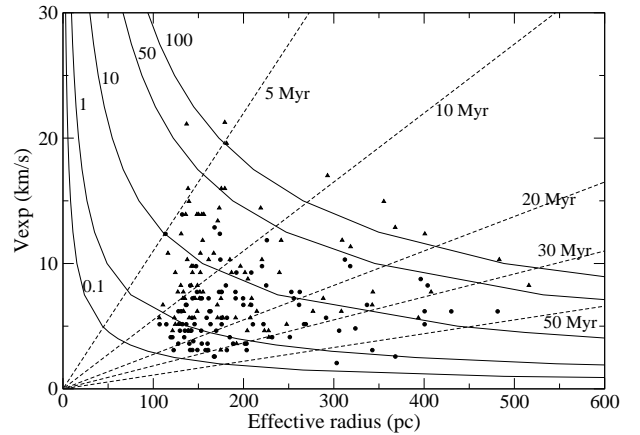}\\
\caption{Effective radius versus expansion velocity. Triangles and circles show the supershell candidates in the second and third Galactic quadrants, respectively. Black dashed straight lines show the curves of constant expansion time. Black solid lines correspond to constant values of $L_w/n$ ($L_w$ is the wind  mechanical luminosity  and $n$ is the interstellar medium density). Black dashed and solid lines were computed using the model of \cite{wea77}.}
\label{reff-velexp}
\end{figure}

The relationship between the effective radius of the supershell candidates and their expansion velocities is shown in Fig. \ref{reff-velexp}. We recall that the expansion velocity of the supershell candidates is calculated as $0.5\, \Delta v$, where  $\Delta v$ is the  velocity range where the structure is detected, and the effective radius is defined as $R_{eff} = \sqrt{\rm a\,\rm b}$, where a and b are the major and minor semi-axes, respectively, of the fitted ellipse. 
Bearing in mind that one of the possible origins  for these structures  may be the cumulative action of the stellar winds for a high number of massive stars, in Fig. \ref{reff-velexp} we have also plotted lines of constant  dynamical age  and constant ratio $L_w/n$, where $L_w$ is the wind  mechanical luminosity  and $n$  the interstellar medium density. Following the analytical  solution of  \cite{wea77}, the dynamical age is  derived from  

$$ t_{dyn}= 0.55 \, \frac{R_{eff}}{v_{exp}} \qquad [\mathrm{Myr}]$$

\noindent where $R_{eff}$ is expressed in units of pc, and $v_{exp}$ is given in  \kms. The constant 0.55 represents a  mean value for the energy and momentum  conserving models. To derive a mean dynamical age for the supershells, we need to know the dominant mechanism at work for their creation. Since this process (or processes) have not been pinned down yet, under the assumption that stellar winds and supernova explosions are the main formation agent, we use a mean value of the constant just for illustrative purposes.

In Fig. \ref{reff-velexp} lines of constant dynamical age (in units of $10^6$ yr) and lines for constant $L_w/n$ (in units of $10^{36}$ erg/s cm$^{-3}$) are shown. Were this mechanism at work for the genesis of most of the structures  listed in our catalog, it would imply that  most of the structures  have dynamical ages between $(5-50) \times 10^ 6$ yr, with only three structures  younger than $5 \times 10^ 6$ yr. This result agrees with the findings of \cite{ehl05,ehl13}.
On the other hand,  most of the structures are between $0.1 \times 10^{36} < \frac{L_w}{n} < 100 \times 10^{36}$ erg/s cm$^{-3}$. Under the assumption of a stellar  origin, such structures cannot be created by the action of only one star. 

It is important to mention that the model of \cite{wea77} does not consider the supernova explosions (SNe), which are the dominant energy source at later evolutionary stages. For illustrative purposes, following \cite{mcc87} and assuming
a density of $n=1$ cm$^{-3}$ and an age $t = 5 \times 10^6$ yr to create a structure having $R_{eff} = 150$ pc, roughly $70$ OB stars with masses $\gtrsim$ 7 M$_\odot$ would be needed, approximately seven of which should have experienced a supernova explosion.

\section{Comparison with other catalogs}

In this section we compare our catalog with the catalogs of \cite{hei79,hei84}, \cite{mcc02}, \cite{ehl05}, and \cite{ehl13}, because they used similar databases with comparable angular resolution to the one used in this work.

It is worth mentioning that only the catalogs of \cite{hei79,hei84} deal with \hi\, supershells. The other catalogs may contain a mixture of structures having small and moderate lineal dimensions (usually called ``shells''), and those having large sizes or requiring a large amount of energy for their creation. The latter  are usually termed ``supershells''.

To carry out the comparison with the above catalogs, we only considered those structures fulfilling our selection criteria. With the aim of identifying the structures that meet our selection criterion \textbf{d)}, we  recalculated the distances and the linear sizes of those structures whose  distances  were obtained with a different galactic rotation model than the one used in this paper. 

We have defined that a simultaneous detection exists between  structures belonging to our catalog and the ones of other catalogs  when  the following conditions are simultaneously met:

\begin{enumerate}

\item The  angular difference of their central coordinates  is lower than the 50\% of the effective radius of the supershell candidate  cataloged in this work.

\item The difference between their systemic velocities is lower than 8 \kms.

\end{enumerate}

A concise summary of the comparison among our catalog and those
 mentioned above is given in Table \ref{coin}. There, the first column 
identifies the catalog we are comparing with, the second gives the total 
number (N$_0$) of structures listed in the catalog, the third one provides the number (N$_1$) of objects of the catalog cited in the first column that meet
 our selection criteria and are located  within the Galactic longitude
range analyzed by us. The fourth column gives the number (N$_2$) of structures
 common to both surveys (the one listed in the first column and ours). The fifth column gives the percentage (\%) of the common detected structures between the different catalogs and ours. Finally,
the last column provides the table number listing the structures common to both
 surveys. In Tables \ref{coinc-hei79}, \ref{coinc-mcc02}, \ref{coinc-ehl05}, and \ref{coinc-ehl13}, the first column always gives the name of the
 \hi\,  supershell candidate as quoted in our catalog. The second and third
 columns
depict the angular extent of the feature in Galactic longitude and Galactic
latitude, and the extent of the velocity range along which the feature is seen,
 in a format like ($\Delta l/2 \times \Delta b/2 \times \Delta v$).
 When performing the comparison with the catalog of \cite{ehl05}, we have
 only considered those features classified by them as having the maximum
 confidence level (index 1 in their nomenclature).

\begin{table*}
 \caption{Structures in common with other catalogs. Notes. N$_0$: total number of the structures listed in each catalog; N$_1$: number of structures of each catalog that fulfills  our selection criteria;  N$_2$: number of structures common between the listed catalogs and ours; and \%: percentage of the common detected structures between the different catalogs and ours.}
\label{coin}
\centering
\begin{tabular}{l r r r r r}
\hline\hline
Catalog&N$_0$&N$_1$&N$_2$&$\%$&Notes \\
\hline
\cite{hei79} & 63 & 14 & 5 & $\sim$36 & Table \ref{coinc-hei79}\\
\cite{hei84} & 42 & 19 & 9 & $\sim$47 & Table \ref{coinc-hei79} \\
\cite{mcc02} & 19 & 4 & 3 & 75 & Table \ref{coinc-mcc02}\\
\cite{ehl05} & 30 & 18 & 11 & $\sim$61 & Table \ref{coinc-ehl05}\\
\cite{ehl13} & 108 & 88 & 34 & $\sim$39 & Table \ref{coinc-ehl13}\\
\hline
\end{tabular}
\end{table*}

\begin{table}
\caption{Structures cataloged by \cite{hei79,hei84} that were detected by our algorithm. * names of the structures as they appear in our catalog.}
 \center
 \label{coinc-hei79}
 \begin{tabular}{l l l}
 \hline\hline
\ Structure* &  Heiles (1979) & Our catalog\\
\ & deg$^2$ \kms & deg$^2$ \kms\\
\hline
\ GS 091$-$04$-$067 &4.5 x 5.0 x 28.0 & 4.6 x 2.7 x 35.0\\
\ GS 094+03$-$110 & 5.0 x 2.5 x 20.0   & 3.2 x 2.5 x 23.4\\
\ GS 094+03$-$120 & 5.0 x 2.5 x 20.0 & 1.9 x 1.2 x 16.5\\
\ GS 108$-$03$-$022& 2.5 x 5.5 x 24.0 & 2.1 x 1.8 x 27.8\\
\ GS 228$-$06+048 & 3.5 x 3.5 x 28.0 &1.8 x 1.3 x 10.3\\
\hline
\  &  Heiles (1984) & \\
\hline
\ GS 096+16$-$025 & 4.0 x 5.5 x 8.9 & 6.4 x 4.5 x 25.8\\
\ GS 105+09$-$021 & 4.5 x 4.0 x 8.0   & 3.9 x 3.2 x 27.8\\
\ GS 124$-$09$-$043 & 3.0 x 4.0 x 12.6 &2.9 x 2.6 x 39.2\\
\ GS 125+11$-$063 & 4.5 x 4.0 x 45.5 & 4.6 x 2.7 x 35.0\\
\ GS 134$-$25$-$020& 8.5 x 8.5 x 8.9 & 7.3 x 6.5 x 17.5\\
\ GS 134+06$-$038 & 5.5 x 3.0 x 8.4 &5.0 x 3.3 x 23.7\\
\ GS 144+08$-$031 &3.5 x 3.0 x 7.4 &4.9 x 4.2 x 18.6\\
\ GS 153+12$-$044 & 5.5 x 3.0 x 8.4 & 3.3 x 2.5 x 23.7\\
\ GS 244$-$16+034 & 4.0 x 3.5 x 8.5 & 3.2 x 2.6 x 16.5 \\
\hline
\end{tabular}
\end{table}

\begin{table*}
 \caption{Structures of the \cite{mcc02} catalog detected by our algorithm. * names of the structures as they appear in our catalog.}
 \center
\label{coinc-mcc02}
 \begin{tabular}{l l l}
 \hline\hline
\ Structure* &  \cite{mcc02} & Our catalog\\
\ & deg$^2$ \kms & deg$^2$ \kms\\
\hline
\ GS 255$-$01+055 & 1.9 x 1.85 x 36  & 1.6 x 1.2 x 13.4\\
\ GS 257+00+067 & 1.1 x 1.1 x 24   & 1.7 x 1.1 x 7.2\\
\ GS 263+00+048 & 1.15 x 1.45 x 26 & 2.2 x 1.5 x 13.4\\
\hline
\end{tabular}
\end{table*}

\begin{table*}
\caption{Structures of the \cite{ehl05} catalog detected by our algorithm. * names of the structures as they appear in our catalog.}
\center
\vspace{0.5cm}
\label{coinc-ehl05}
\begin{tabular}{l l l}
\hline\hline
\ Structure*&\cite{ehl05}&Our catalog\\
\ &deg$^2$ \kms &deg$^2$ \kms\\
\hline
\ GS 093$-$14$-$021 & 7.0 x 5.0 x 18.5 & 5.0 x 5.0 x 22.7\\
\ GS 96+16$-$025 & 4.2 x 6.0 x 15.5 & 6.4 x 4.5 x 25.8\\
\ GS 103+07$-018$ & 11.0 x 4.7 x 23.7 & 6.9 x 3.9 x 20.6 \\
\ GS 107+13$-$040 &  1.5  x  1.2 x 25.8 & 2.9 x 2.2 x 32.0\\
\ GS 110$-$04$-$067 & 1.7 x  1.7 x 23.7 & 2.3 x 1.9 x 32.0\\
\ GS 117$-$02$-$121 & 4.0 x 2.5 x 11.3 & 3.6 x 2.4 x 23.7\\
\ GS 128+01$-$103 & 5.5 x 2.2 x 15.5 & 3.2 x 2.0 x 9.3\\
\ GS 130$-$17$-$048 & 6.0 x 7.5 x 15.5 & 7.2 x 4.9 x 16.5\\
\ GS 146+02$-$056 & 4.7 x 3.2 x 18.5 & 1.3 x 0.9 x 8.2\\
\ GS 218$-$05+037 & 7.0 x 3.7 x 16.5 & 3.1 x 2.2 x 9.6\\
\ GS 242+05+058 & 2.0 x 2.2 x 21.6 & 2.6 x 2.3 x 8.2\\
\hline
\end{tabular}
\end{table*}

\begin{table}
\caption{Structures of \cite{ehl13} catalog detected by our algorithm. * names of the structures as they appear in our catalog.}
 \center
 \vspace{0.5cm}
 \label{coinc-ehl13}
 \begin{tabular}{l l l}
 \hline\hline
\ Structure* &  \cite{ehl13} & Our catalog\\
\ & deg$^2$ \kms & deg$^2$ \kms\\
\hline
\ GS 093+03$-$031 & 3.2 x 3.2 x 12.4 & 1.9 x 1.3 x 18.6\\
\ GS 093$-$14$-$021 & 6.2 x 4.2  x 35& 5.5 x 5.0 x 22.7 \\
\ GS 098+03$-$115 & 7.2 x 2.5 x 20.6& 2.7 x 2.0  x 27.8\\
\ GS 101$-$02$-$037 & 2.0 x 1.2 x 11.3& 2.4 x 1.8  x 18.6\\
\ GS 103+07$-$018 & 10.7 x 4.5 x 34& 6.9 x 3.9 x 20.6\\
\ GS 105$-$03$-$061 & 4.0 x 1.7 x 8.2& 3.3 x 2.7 x 16.5\\
\ GS 109+06$-$032 & 2.2 x 1.5 x 13.4& 3.0 x 2.1 x 20.6\\
\ GS 109$-$08$-$065 & 1.7 x 1.2 x 12.4& 2.8 x 1.8 x 17.5\\
\ GS 110$-$04$-$067 & 1.5 x 1.2 x 9.3& 2.3 x 1.9 x 32.0\\
\ GS 111+08$-$041 & 1.7 x 1.5 x 14.5& 2.4 x 1.7 x 42.3\\
\ GS 112+01$-$102 & 1.0 x 0.7 x 9.2& 1.8 x 1.5 x 15.5\\
\ GS 121$-$05$-$037 & 4.5 x 6.0 x 39.1& 3.8 x 2.6 x 14.4\\
\ GS 129+01$-$108 & 5.2 x 2.0 x 16.5& 2.0 x 1.4 x 9.3\\
\ GS 129+05$-$061 & 3.2 x 1.2 x 20.6& 4.0 x 2.3 x 18.6\\
\ GS 130+00$-$101 & 5.2 x 2.0 x 16.5& 2.9 x 2.0 x 7.2\\
\ GS 136$-$04$-$077 & 8.0 x 2.7 x 21.6& 2.5 x 2.2 x 10.3\\
\ GS 136$-$09$-$033 & 11.5 x 5.2 x 52.6& 4.4 x 2.8 x 11.3\\
\ GS 137+04$-$071 & 1.7 x 0.7 x 8.3& 2.1 x 1.7 x 9.3\\
\ GS 137+06$-$029 & 3.0 x 2.0 x 17.5& 3.5 x 2.7 x 15.5\\
\ GS 144+08$-$031 &  5.0 x 3.0 x 30.9  & 4.9 x 4.2 x 18.6\\ 
\ GS 145$-$09$-$066 & 1.7 x 1.0 x 8.2& 3.3 x 2.0 x 12.4\\
\ GS 148$-$09$-$038 & 2.0 x 1.0 x 8.3& 2.6 x 2.1 x 18.6\\
\ GS 156$-$05$-$061 & 2.7 x 1.0 x 10.3& 2.9 x 1.8 x 13.4\\
\ GS 160+05$-$043 & 2.5 x 2.7 x 11.3& 4.5 x 2.7 x 10.3\\
\ GS 164+00$-$021 & 4.5 x 1.7 x 11.4& 2.1 x 1.8 x 9.3\\
\ GS 198$-$18+26 & 3.5 x 2.7 x 14.4 & 4.1 x 3.0 x 12.4\\
\ GS 229+02+063 & 2.0 x 1.0 x 10.3& 2.3 x 1.4 x 11.3\\
\ GS 239$-$02+068 & 1.2 x 1.0 x 7.2& 1.9 x 1.5 x 14.4\\
\ GS 251$-$08+054 & 1.2 x 1.2 x 10.3& 2.2 x 1.7 x 15.5\\
\ GS 255$-$01+55 & 3.0 x 1.5 x 20.6& 1.6 x 1.2 x 13.4\\
\ GS 257+09+037 & 8.2 x 4.7 x 59.8& 3.7 x 2.8 x 23.7\\
\ GS 265$-$06+082 & 2.5 x 1.5 x 19.6& 2.6 x 2.2 x 11.3\\
\ GS 266$-$05+096 & 2.0 x 2.7 x 11.3& 2.6 x 2.0 x 13.4\\
\ GS 267$-$03+117 & 2.7 x 2.0 x 15.5& 2.7 x 1.6 x 5.2\\
\hline
\end{tabular}
\end{table}

After carrying out the comparisons, we found that the lack of detection in our catalog of a given feature listed in other catalog may be due to a
 series of facts that can be summarized as follows:

\begin{itemize}
 \item {\it Case A}: The structure does not have a clearly defined \hi\,
 central minimum.\\
 In this case all non-detected structures show scatter \hi\, emission
peaks towards their central areas. In this case our algorithm may detect features
 smaller than 2$^\circ$. In other cases it may detect \hi\, maxima
that though at first ``glance'' they look like they belong to an \hi\, wall,
in the end they are too dispersed in position along the radial lines, and do not
follow the expected pattern for a shell-like feature.\\

\item {\it Case B}: The wall of the detected structure has a low brightness temperature along most of its
perimeter.\\
In this case, the \hi\, maxima defining the supershell wall have a peak 
brightness temperature lower than the T$_{rms}$({\it l, b, v}) threshold set
by our algorithm.\\

\item {\it Case C}: The structures being compared have a good positional 
coincidence, but the difference in their systematic radial velocities is
greater than 8 \kms.\\

\item {\it Case D}: The structures being compared have a good positional
 correspondence, but the angular diameter of the structure detected by our 
algorithm is smaller than 2$^\circ$.\\

\item {\it Case E}: The structure is detected by our algorithm, but it is discarded in the final visual revision step.

\end{itemize} 
    
In  Table \ref{no-coin} we give a summary of the causes that prevent
 the identification by our procedure, of structures already listed in existing 
catalogs. It is important to mention that from this analysis we  can conclude that the catalogs  strongly depend on the search methods and on the selection criteria established by  different authors.

\begin{table*}
 \caption{Undetected structures in our catalog.}
 \center
\label{no-coin}
 \begin{tabular}{l r r r r r r}
 \hline\hline
\ Catalog & Case A & Case B & Case C & Case D & Case E & Total \\
\hline
\cite{hei79}, \cite{hei84} & 15 & 3 & 1 & -- & -- & 19\\
\cite{mcc02} & -- & -- & 1 & -- & -- & 1\\
\cite{ehl05} & 3 & 4 & -- & -- & -- & 7\\
\cite{ehl13} & 3 & 43 & 1 & 7 & -- & 54\\
\hline
\end{tabular}
\end{table*}

\section{Conclusions}

 A new catalog of \hi\, supershell candidates has been  constructed
 using a combination of an automatic detection algorithm plus a visual one.
It is known that pure visual identification methods are difficult and very time
 consuming mainly because the dynamic range has to be adjusted quite often to
 make structures visible. However, the eye is an incredibly powerful instrument,
 especially when images are irregular, since it is able to combine disconnected
 patterns. Our algorithm was trained on an initial  visual catalog and then applied
 to the whole dataset. At the end of this process, all the detected structures 
were again carefully inspected by eye  to do the final selection.
In other words,
 in our method, we have combined the power of visual inspection, together 
with the power of a computer-based algorithm working in a supervised (semi-automatic)  mode to optimize the time required to analyze a huge amount of data. 
A total of 566 candidate structures has been detected,  347 in the
second Galactic quadrant and 219 in the third one.

From the distribution of the detected structures in the sky, it can be seen that for the second Galactic quadrant, we have detected structures with distances up to 32 kpc form the Sun, while for the third quadrant all the structures  have distances less than 17 kpc from the Sun. The explanation of this effect is far from clear. It shows that the outer part of the Galaxy deserves a thorough study.

Owing to our selection criteria, the catalog suffers from  selection 
effects.  Bearing this in mind only features closer to the Sun than 5.7 kpc were used to derive their statistical properties.

The estimated mean weighted effective radius is about 160 pc for both Galactic quadrants. The derived eccentricities indicate that about $98 \, \%$ of the supershells are elliptical. The mean weighted eccentricity is 0.8 $\pm$ 0.1  for both Galactic quadrants.

An inspection of the orientation angle values (e.g., the angle between the Galactic plane and the feature's major semi-axis) shows that most of the supershell candidates are elongated parallel to the Galactic plane, in agreement with the conclusions of other researchers.           
Based on this finding, it is believed that the galactic density gradient plays a minor, if any, role in the time evolution of these structures. On the other hand, the magnetic field running parallel to the Galactic disk could be the responsible for the observed effect. Otherwise, the major axes
 of the structures should be predominantly perpendicular to the Galactic 
plane.

The distribution of the centroid of the structures relative to the
 Galactic plane shows that roughly $\sim$ 40\% of the supershells are confined
 to $z \leq 500 $ pc. 
After applying correction factors owing to the incompleteness of our sample, we
 derived the surface density of \hi\, supershells. Though it 
decreases as the galactocentric distance increases, in close agreement with
 the findings of other researchers, the actual surface density in the solar
neighborhood is almost a factor of 2 higher, 7.4  kpc$^{-2}$, than the one derived before by
 \cite{ehl13}. This could  be a direct consequence of the
``ability'' of our method to identify incomplete features. In line with this,
we would like to point out that roughly only half of the supershell candidates are found
 to be completely closed. The decrease in  the surface density with galactocentric 
distance is well fit by a Gaussian function with a radial scale length of 
 4.4 $\pm$ 0.3 kpc. 
 The surface (f$_{2D}$) and volume (f$_{3D}$) filling factors are f$_{2D} = 0.5 \pm 0.1$ and f$_{3D} = 0.04^{+0.01}_{-0.02}$.
 As mentioned above, because our algorithm is able to detect incomplete structures, we have, as a byproduct, a catalog of ``galactic chimney'' candidates, which contains  80 structures.

A clear  relationship between the effective radius of the
structures and their expansion velocities is not detected in our catalog.
Under the assumption  that their genesis is mostly consequence of the action 
 of massive stars (stellar winds and supernova explosions), the dynamic age of the structures are between 5-50 Myr and most of the
 structures fall between values of $L_w/n$ of (0.1-100) $\times 10^{36}$ erg/s cm$^{-3}$.

A comparison with the structures listed in our catalog with
those given in other shell/supershell catalogs, shows that we identify, on average, 57\% of the structures listed elsewhere.
 The lowest correspondence is
with the catalog of \cite{hei79} ($\sim$ 36\%), and the highest is with 
\cite{mcc02} ($\sim$ 75 \%).

\section {Acknowledgments}

This work was partially supported by the grants PIP01299/2008 and 11/G091 of
the Consejo Nacional de Investigaciones Cient\'{\i}ficas y T\'ecnicas (CONICET)
of Argentina, and Universidad Nacional de La Plata (UNLP), respectively.
We acknowledge Dr. Ariel Chernomoretz for having produced  Figure 3.
We would like to thank an anonymous referee for  constructive and useful comments that helped us to considerably improve the quality of  the original  paper.

\bibliographystyle{aa} 
\bibliography{bibliografia}
  
 \IfFileExists{\jobname.bbl}{}
{\typeout{}
\typeout{****************************************************}
\typeout{****************************************************}
\typeout{** Please run "bibtex \jobname' ' to optain}
\typeout{** the bibliography and then re-run LaTeX}
\typeout{** twice to fix the references!}
\typeout{****************************************************}
\typeout{****************************************************}
\typeout{}
}

\end{document}